\documentclass[showpacs,nofootinbib,preprintnumbers,prd,superscriptaddress,twocolumn]{revtex4}

\usepackage{amsmath}
\usepackage{braket}
\usepackage{amsfonts}
\usepackage{amssymb}
\usepackage{graphicx}
\usepackage[titletoc]{appendix}
\usepackage{color}
\usepackage{hyperref}
\usepackage{cleveref}
\usepackage[rightcaption]{sidecap}
\usepackage{subfigure}

\usepackage{float}
\usepackage{wasysym}
\usepackage{amssymb}
\usepackage{tensor}
\usepackage{csquotes}

\usepackage{dcolumn}

\usepackage{array}
\usepackage{ctable}
\usepackage{multirow}
\usepackage{siunitx}
\usepackage{longtable}
\usepackage{tabularx}
\usepackage{booktabs}

\usepackage{mathtools}
\usepackage{dirtytalk}
\usepackage{siunitx}
\DeclareSIUnit{\parsec}{pc}

\graphicspath{{Graphics/}}

\def\be{\begin{equation}}
\def\ee{\end{equation}}
\def\bea{\begin{eqnarray}}
\def\eea{\end{eqnarray}}

\newcommand{\virgolette}[1]{``#1''}

\definecolor{vividviolet}{rgb}{0.62, 0.0, 1.0}
\definecolor{amaranth}{rgb}{0.9, 0.17, 0.31}
\definecolor{palatinateblue}{rgb}{0.15, 0.23, 0.89}
\definecolor{brightpink}{rgb}{1.0, 0.0, 0.5}
\definecolor{cornflowerblue}{rgb}{0.39, 0.58, 0.93}
\definecolor{deepcarminepink}{rgb}{0.94, 0.19, 0.22}
\definecolor{radicalred}{rgb}{1.0, 0.21, 0.37}

\hypersetup{ linktoc=all,
    colorlinks, linkcolor={palatinateblue},
    citecolor={brightpink}, urlcolor={amaranth}
}

\begin{document}

\title{Entanglement entropy evolution during gravitational collapse}

\author{Alessio Belfiglio}
\email{alessio.belfiglio@unisi.it}
\affiliation{DSFTA, University of Siena, Via Roma 56, 53100 Siena, Italy.}

\author{Orlando Luongo}
\email{orlando.luongo@unicam.it}
\affiliation{Istituto Nazionale di Fisica Nucleare (INFN), Sezione di Perugia, Perugia, 06123, Italy.}
\affiliation{School of Science and Technology, University of Camerino, Via Madonna delle Carceri, 62032, Italy.}
\affiliation{Istituto Nazionale di Astrofisica (INAF), Osservatorio Astronomico di Brera, Milano, Italy.}
\affiliation{Al-Farabi Kazakh National University, Almaty, 050040, Kazakhstan.}

\author{Stefano Mancini}
\email{stefano.mancini@unicam.it}
\affiliation{Istituto Nazionale di Fisica Nucleare (INFN), Sezione di Perugia, Perugia, 06123, Italy.}
\affiliation{School of Science and Technology, University of Camerino, Via Madonna delle Carceri, 62032, Italy.}

\author{Sebastiano Tomasi}
\email{sebastiano.tomasi@unicam.it}
\affiliation{School of Science and Technology, University of Camerino, Via Madonna delle Carceri, 62032, Italy.}
\affiliation{Istituto Nazionale di Fisica Nucleare (INFN), Sezione di Perugia, Perugia, 06123, Italy.}

\date{\today}

\begin{abstract}
We investigate the dynamics of the ground state entanglement entropy for a discretized scalar field propagating within the Oppenheimer-Snyder collapse metric. Starting from a well-controlled initial configuration, we follow the system as it evolves toward the formation of a horizon and, eventually, a singularity. Our approach employs an Ermakov-like equation to determine the time-dependent ground state of the field and calculates the resulting entanglement entropy by tracing out the degrees of freedom inside a spherical region within the matter sphere. We find that the entanglement entropy exhibits nontrivial scaling and time dependence during collapse. Close to the horizon, the entropy can deviate from the simple area law, reflecting the rapid changes in geometry and field configuration. Although the model is idealized, these results provide insights into the generation and scaling of entanglement in the presence of realistic, dynamically evolving gravitational fields.
\end{abstract}



\maketitle
\tableofcontents

\section{Introduction}

In classical thermodynamics, entropy is typically extensive. However,  dealing with quantum mechanical systems, it is possible to argue an entropy that can be present even if the state of the entire system is known. In particular, a nonzero von Neumann entropy may generally arise for discretized quantum fields if the overall degrees of freedom are divided into two distinct subsets, and the corresponding ground-state reduced density operators are derived.
Accordingly, the associated entropy is named \emph{entanglement entropy} and it quantifies the field quantum correlations, provided the state of the full system is pure \cite{qit,pl1}.

Early computations involving a free scalar field in flat spacetime \cite{qft1} revealed that the resulting von Neumann entropy grows in proportion to the boundary area of the region which is traced out. This result was later confirmed in the case of a spherical entangling surface by determining the reduced density matrix for the scalar field modes inside the sphere \cite{qft2}. The emergence of such \emph{area law} \cite{rev1}, typically up to subleading logarithmic corrections, is closely related to the locality of interactions in quantum many-body systems \cite{PhysRevLett.93.227205,PhysRevLett.94.060503,PhysRevB.73.085115,PhysRevLett.96.220601,RevModPhys.80.517} and it manifests striking similarity with Bekestein-Hawking black hole entropy \cite{bh1, bh2, bh3, bh4, Wald:1999vt, Page:2004xp}. This resemblance has then fueled the idea that black hole entropy might be fundamentally entanglement-based, potentially connecting quantum information tools to quantum field theory \cite{PasqualeCalabrese_2004, Casini_2009, RevModPhys.90.045003} and to the holographic principle \cite{holo, PhysRevLett.96.181602, Nishioka:2009un, Rangamani:2016dms, RevModPhys.90.035007}. Key developments in this direction \cite{PhysRevD.73.121701, Das_2007} established that the area law is preserved for discretized scalar fields in static, spherically symmetric spacetimes, thus strengthening the link between horizon geometry and entanglement \cite{Das_2007,Solodukhin:2011gn}. Accordingly, the scalar field’s ground state may effectively encode horizon-related degrees of freedom, supporting the idea that black hole entropy can be interpreted in terms of quantum correlations across the horizon\footnote{This perspective can be also inferred by artificially introducing a horizon in flat spacetime, see, e.g., \cite{Das_2007} for further discussions.}. Possible connections between field entanglement and black hole entropy have been recently generalized to other thermodynamic quantities, suggesting a broader correspondence between entanglement “mechanics” and black hole thermodynamics \cite{PhysRevD.102.125025}. However, when relating black hole properties to quantum field entanglement, it might be noted that the field entropy is sensitive to the ultraviolet (UV) cutoff introduced within the discretization procedure, while the Bekenstein-Hawking entropy is finite by definition \cite{Solodukhin:2011gn}.

Such UV sensitivity can be further investigated through field smearing in disjoint spatial regions \cite{PhysRevD.104.085012, Martin:2021qkg} or by imposing a UV cutoff close to Planck lengthscales, which may allow to determine possible corrections to entanglement entropy in effective quantum gravity theories \cite{Belfiglio:2024qsa}, with the ultimate aim of interpreting such entropy as a microscopic, quantum-gravitational contribution to black hole entropy. At the same time, possible area law violations arising from nonminimal field-curvature coupling have been recently studied \cite{Belfiglio:2023sru}, aiming to provide further insights into the effects of geometry on the entanglement entropy scaling in discrete field theories.

Although substantial progress has been made in static scenarios, the question of how entanglement entropy behaves in position space under fully dynamical gravitational conditions remains largely open\footnote{On the other hand, momentum-space entanglement \cite{PhysRevD.86.045014} has been widely studied in dynamical cosmological settings, with particular attention to interacting field theories in early-time scenarios \cite{PhysRevD.102.043529,Brahma:2021mng, PhysRevD.105.123523, PhysRevD.107.103512,PhysRevD.108.043522, PhysRevD.109.123520}.}. Initial studies performed in $(1+1)$-dimensional Friedmann-Lemaître-Robertson-Walker (FLRW) spacetimes \cite{Boutivas:2023ksg,Katsinis:2023hqn} were subsequently generalized to realistic cosmological models, mainly focusing on discrete scalar fields, in the attempt to mimic the evolution of scalar inflationary perturbations up to the radiation-dominated era and to understand their quantum-to-classical transition \cite{PhysRevD.77.063534,PhysRevD.109.023503,Boutivas:2023mfg}.

However, a proper characterization of real-space entanglement within dynamical black hole scenarios is still missing, despite it would inevitably provide fundamental insights into the quantum properties of black holes and the origin of their entropy.

A simple yet instructive model for gravitational collapse is represented by the Oppenheimer-Snyder (OS) solution \cite{PhysRev.56.455,Weinberg1972,Malafarina2017}, which encodes essential features of black hole formation by describing a uniform dust cloud collapsing into a Schwarzschild black hole. Inside the dust cloud, the metric is given by a closed FLRW solution, which smoothly matches the Schwarzschild exterior. The OS model’s simplicity and tractability make it a suitable arena to explore how entanglement entropy behaves as horizons form and evolve dynamically. This setting then allows us to probe the robustness of the area law scaling under conditions where geometry and matter fields may evolve in time.

In order to provide a direct comparison with previous investigations, in this work we focus on the scaling behavior of the ground state entanglement entropy for a discretized scalar field during OS collapse. The ground state dynamics is governed by Ermakov-like equations, which we solve in order to study the time evolution of the corresponding entanglement entropy, as the system approaches horizon formation. Tracing out a fixed number of degrees of freedom within the collapsing region, we show that spatial curvature effects modify the entropy scaling during collapse, obtaining significant deviations from area law in the presence of a sufficiently large curvature parameter. We also observe that the largest contribution to the total entropy is produced during the latest stages of collapse, while such entropy may even decrease in time during the early phases. We finally confirm that the total amount of entanglement is observer-dependent, highlighting the main differences between a comoving description of collapse and the expected corrections for a Schwarzschild observer at spatial infinity. Despite the OS model has a purely classical origin\footnote{See, for example, Refs. \cite{PhysRevD.101.026016,PhysRevLett.130.101501,QOS_plb} for some proposals of quantum gravitational collapse and the formulation of a quantum OS model.}, our results may represent
a first step towards more general investigations of non-static horizons in semiclassical models and quantum gravity proposals, where singularities may be resolved.

The paper is organized as follows. In Sec.~\ref{sec:OS}, we review the OS collapse model. In Sec.~\ref{sec:theoreticalsetup}, we discuss our theoretical setup for discretizing the field theory and computing entanglement entropy. Our main results, including entropy scaling and its time dependence, are presented in Sec.~\ref{sec:results}. Finally, in Sec.~\ref{sec:conclusions}, we summarize our findings and suggest potential avenues for future research. Throughout this paper, we will consistently use natural units $c=\hbar=1$.

\section{Gravitational collapse model}\label{sec:OS}

The OS collapse solution \cite{Weinberg1972,Malafarina2017} is a general relativistic model describing the collapse of an initially static, spherically symmetric and pressureless matter cloud. It can be straightforwardly derived from general relativity, providing Friedmann-like equations for the dynamics of the collapsing object. One finds that the interior of the spherical collapsing region is described by a FLRW metric\footnote{By identifying the radius of the matter sphere $r_b a(t)=R(t)$, with the scale factor satisfying $a(t_0)=1$ and $t_0$ denoting the initial time of collapse.} that possesses a non-zero spatial curvature,
\begin{equation}\label{eq:close_flrw_metric}
    \mathrm{d}s^2=\mathrm{d}t^2 - a^2(t)\left(\frac{\mathrm{d}r^2}{1-kr^2}+r^2\mathrm{d}\Omega^2\right),
\end{equation}
where $d\Omega^2=d\theta^2+\sin^2\theta d\phi^2$ and $k$ given by
\begin{equation}\label{eq:spatial_curvature}
    k=\frac{2GM}{r_b^3}=\frac{r_s}{r_b^3},
\end{equation}
with $r_s$ denoting the Schwarzschild radius corresponding to the total mass $M$.

The solution for the scale factor as a function of time is usually given in parametric form\footnote{A simple expression can be obtained for $\alpha \rightarrow 0$, $a(t)\approx (9/4 k)^{1/3}(t-t_c)^{2/3}$ where $t_c$ is the collapse time. By introducing $\bar{t}=t-t_c$, we notice that $a(t)\propto \bar{t}^{2/3}$. Thus, near the collapse, the OS model can be approximated by a collapsing EdS universe.}
\begin{equation}\label{eq:parametric_solution}
    a(\alpha)=\frac{1+\cos(\alpha)}{2},\qquad t(\alpha)=\frac{1}{\sqrt{k}}\frac{\alpha+\sin(\alpha)}{2},
\end{equation}
where $\alpha=0$ corresponds to the initial time $t=0$. The collapse continues until the radius of the matter sphere reaches zero at $\alpha=\pi$, so the collapse starts at $t=0$ and ends at $t_c=\frac{\pi}{2\sqrt{k}}$. For a comoving observer, such collapse then occurs in a finite proper time.

The time required to reach the Schwarzschild radius $r_s$ is found by solving $a(\alpha_{r_s})r_b=r_s$ and substituting $\alpha_{r_s}$ into the parametric solution:
\begin{align*}
t_{r_s} = t(\alpha_{r_s})= \frac{1}{2 \sqrt{k}} \bigg[ 2 \sqrt{\, \frac{r_s}{r_b} - \frac{r_s^2}{r_b^2} \,} +\, \arccos\left( \frac{2 r_s}{r_b} - 1 \right) \bigg],
\end{align*}
with $a(\alpha_{r_s})=r_s/r_b=kr_b^2.$

Next, we consider the perspective of an external observer, which requires to determine the metric outside the collapsing object. The assumption of spherical symmetry then implies that the external geometry is described by the Schwarzschild solution
\begin{equation}\label{eq:schwarzschild_metric}
    \mathrm{d}s^2_{\text{ext}} = \left(1 - \frac{r_s}{R}\right)\mathrm{d}T^2 - \frac{1}{\left(1 - \frac{r_s}{R}\right)}\,\mathrm{d}R^2 - R^2\,\mathrm{d}\Omega^2,
\end{equation}
where the variables $T$ and $R$ are the external temporal and radial coordinates, respectively. The internal and external metrics are then matched at the matter boundary $r_b$, to ensure a continuous global collapse metric and thus avoiding nonphysical energy shells.
Setting the boundary as $R=R_b(T)$ in the Schwarzschild exterior and $r=r_b$ in the FLRW interior, both metrics induce a $2+1$-dimensional metric on the boundary. Equating these induced metrics ensures the continuity of the solution, providing the proper Darmois-Israel junction conditions \cite{Bonnor1981}.
The conditions for continuity in comoving time read
\begin{align}\label{eq:dT_dt}
    &\frac{\mathrm{d}T}{\mathrm{d}t} = \frac{\sqrt{1 - \frac{r_s}{R_b(t)} + \left(\frac{\mathrm{d}R_b}{\mathrm{d}t}\right)^2}}{1 - \frac{r_s}{R_b(t)}},\\
    &R_b(t) = a(t)\, r_b,
\end{align}
while the continuity of the extrinsic curvature follows directly from these conditions.

The exterior can be computed starting from the metric in Novikov coordinates and replacing $r$ by $r^*$, then performing the matching. See for example \cite{misner1973gravitation}. The global spacetime metric can then be expressed in comoving coordinates as
\begin{equation}\label{eq:global_metric_comov}
    \mathrm{d}s^2 = \mathrm{d}t^2 - a^2(t) \left( \frac{\mathrm{d}r^2}{1 - \frac{r_s}{r_b^3}r_{-}^2 r_{+}^{-1}} + r^2 \mathrm{d}\Omega^2 \right),
\end{equation}
where $ r_{-} = \min(r, r_b) $, $ r_{+} = \max(r, r_b) $, and $ a(\alpha) $, $ t(\alpha) $ are given by Eq.~\eqref{eq:parametric_solution}, with $ k $ replaced by $ k_{+} = r_s/r_{+}^3 $.

For $r \geq r_b$, Eq.~(\ref{eq:global_metric_comov}) reduces to the Schwarzschild metric in Novikov coordinates \cite{misner1973gravitation}, representing the exterior solution, while for $r < r_b$ it reduces to Eq.~(\ref{eq:close_flrw_metric}) \cite{Weinberg1972}.

The $t$ coordinate represents the proper time measured by comoving observers who move radially inward along with the collapsing matter inside the cloud. Outside the matter distribution, $t$ serves as well as the proper time for observers who free-fall radially from rest at some finite initial altitude in the Schwarzschild geometry. In other words, all observers described as comoving in this coordinate system experience $t$ as their own proper time, whether they start deep within the collapsing matter or begin falling from the vacuum exterior region.

As the collapse proceeds, a curvature singularity forms at $r=0$. Every comoving observer who is initially inside the collapsing object reaches this singularity at the same coordinate time $t$, reflecting the simultaneous nature of the end of their worldlines. Outside the matter, observers starting from higher altitudes fall inward more gradually, reaching the singularity at later coordinate times, thus highlighting that $t$ increases with altitude.

\section{Ground state of the discretized field}\label{sec:theoreticalsetup}
The action of a massive scalar field propagating in a spatially curved FLRW spacetime can be expressed in the form
\begin{equation}\label{eq:action_sf_flrw}
    \begin{aligned}
        2S &= \int_{\mathcal{D}} \mathrm{d}^4x\sqrt{-g}\mathcal{L} \\
           &= \int_{\mathcal{D}} \mathrm{d}^4x\sqrt{-g}\left\{g^{\mu\nu}\partial_\mu\phi\partial_\nu\phi-\mu^2\phi^2\right\} \\
           &=\int_{\mathcal{D}} \mathrm{d}t\,\frac{r^2\,\mathrm{d}r}{\sqrt{1-kr^2}}\mathrm{d}\Omega \, a^3(t)\bigg(\dot{\phi}^2\\&-\frac{1}{a^2(t)}\bigg[(1-kr^2)|\nabla\phi|^2_{\text{rad}}+|\nabla\phi|^2_{\text{ang}} \bigg] -\mu^2\phi^2 \bigg),
    \end{aligned}
\end{equation}
where $\mu$ is the field mass, $k$ is the spatial curvature, $a(t)$ is the scale factor in terms of the comoving time $t$ and $g$ is the determinant of the metric, while $\mathcal{D}$ is the integration domain. The radial and angular parts of the gradient are implicitly defined.

The curved space Lagrangian is defined by the relation $ S=\int_{\mathcal{D}}L(t)\mathrm{d}t$, where $L(t)=\int_{\mathcal{D}} \mathrm{d}^3x\sqrt{-g}\mathcal{L}$. Thus, we readily find
\begin{equation}\label{eq:sf_FLRW_spacetime_Lagrangian}
    \begin{aligned}
        2L&=\int_{\mathcal{D}}\frac{r^2\,\mathrm{d}r}{\sqrt{1-kr^2}}\mathrm{d}\Omega\, a^3(t)\bigg(\dot{\phi}^2\\
        &-\frac{1}{a^2(t)}\bigg[(1-kr^2)|\nabla\phi|^2_{\text{rad}}+|\nabla\phi|^2_{\text{ang}} \bigg] -\mu^2\phi^2 \bigg).
    \end{aligned}
\end{equation}


\subsection{Discretization of the scalar field theory}

The scalar field Lagrangian in Eq.~(\ref{eq:sf_FLRW_spacetime_Lagrangian}) is discretized using the procedure outlined in Appendix \ref{appendix:sph_harm_discretization_FLRW}. The corresponding Hamiltonian is given by
\begin{equation}\label{eq:discretized_ham}
    \begin{aligned}
        2H &= K + V=\\
           &= \sum_{lmj} \left[ \frac{\pi^2_{lmj}}{a^3}  +\, a\left(j+\frac{1}{2}\right)^2\sqrt{1 - k b^2\left(j+\frac{1}{2}\right)^2} \right. \\
           &\quad \left. \times \left[\frac{\left(1 - k b^2 (j+1)^2\right)^{1/4}}{j+1}\phi_{lm,j+1} - \frac{\left(1 - k b^2 j^2\right)^{1/4}}{j}\phi_{lmj} \right]^2 \right. \\
           &\quad \left. +\, a\left( \frac{l(l+1)}{j^2} + (\mu a b)^2 \right) \phi_{lmj}^2 \right],
    \end{aligned}
\end{equation}
where the cutoff $b$ is absorbed through appropriate canonical transformations \cite{chandran2023dynamical}. This transformation implies a rescaling of the time coordinate by the cutoff, thus providing a dimensionless time variable. In Eq.~(\ref{eq:discretized_ham}), we display the kinetic and potential components of the discretized Hamiltonian. The kinetic term is given by $ K = \sum_{lmj} \pi^2_{lmj}/a^3 $, and the remaining sum constitutes the potential term. Focusing on the potential term in Eq.~(\ref{eq:discretized_ham}), after some algebraic manipulation, we obtain

\begin{widetext}
\begin{equation}\label{eq:potential_expanded}
    \begin{aligned}
    V = \sum_{j=1}^{N} \bigg\{ &\frac{(1+2j)^2}{4 (j+1)^2}\sqrt{[1 - k b^2 (j+1)^2][1 - k b^2 (j + \frac{1}{2})^2]}\,\phi_{lm,j+1}^2 \\
    &- 2\frac{(1+2j)^2}{4 j (j+1)} \left[(1 - k b^2 j^2)(1 - k b^2 (j+1)^2)\right]^{1/4} \sqrt{1 - k b^2 (j + \frac{1}{2})^2}\,\phi_{lmj} \phi_{lm,j+1} \\
    &+ \left[\sqrt{(1 - k b^2 j^2)(1 - k b^2 (j + \frac{1}{2})^2)}\left(1 + \frac{1}{j} + \frac{1}{8j^2}\right) + \frac{l(l+1)}{j^2} + (\mu a b)^2 \right]\,\phi_{lmj}^2 \bigg\},
    \end{aligned}
\end{equation}
\end{widetext}

where we omit for simplicity the dependence on $l$ and $m$. Further, we rewrite the potential in Eq.~(\ref{eq:potential_expanded}) by introducing parameters $ A_j $, $ B_j $, and $ D_j $:
\begin{equation}\label{eq:quadratic_form}
    V = \sum_{j=1}^{N} \left( A_j \phi_{j+1}^2 - 2 B_j \phi_{j+1} \phi_{j} + D_j \phi_{j}^2 \right).
\end{equation}
We now define the coupling matrix $\tilde{\boldsymbol{C}}$, which determines the quadratic form in Eq.~(\ref{eq:quadratic_form}). It is given by
\begin{equation}\label{eq:coupling_matrix_general}
    \begin{aligned}
        &\tilde{\boldsymbol{C}}_{11}=D_1,\\
        &\tilde{\boldsymbol{C}}_{ii}= D_i+A_{i-1},\quad i\neq N,1\\
        &\tilde{\boldsymbol{C}}_{i,i-1}=\tilde{\boldsymbol{C}}_{i-1,i}=-B_i,\quad \\
        &\tilde{\boldsymbol{C}}_{NN}=A_{N-1}.
    \end{aligned}
\end{equation}
The coupling matrix acts on $\boldsymbol{\phi}_{lm}=(\phi_{lm1},\ldots,\phi_{lmN})^T$. In the Hamiltonian, it appears as $a(t)\tilde{\boldsymbol{C}}$. By substituting the parameters  $ A_j $, $ B_j $, and $ D_j $ into Eq.~(\ref{eq:coupling_matrix_general}), simplifying, and decomposing $\tilde{\boldsymbol{C}}$ into a time-independent component $\boldsymbol{C}$ and a time-dependent component $\boldsymbol{T}(t)$, we obtain $\tilde{\boldsymbol{C}} = \boldsymbol{C} + \boldsymbol{T}(t)$, where
\begin{equation}\label{eq:coupling_matrix}
\begin{aligned}
    &\begin{cases}
        \boldsymbol{C}_{11}=\frac{9}{4}\sqrt{(1-\frac{9}{4}kb^2)(1-kb^2)}+l(l+1),\\[1.8ex]
        \boldsymbol{C}_{ii}= \frac{ \sqrt{1-kb^2 j^2 }}{ j^2}\left( ( j+1/2)^2\sqrt{1-kb^2 \left(j+1/2\right)^2 }\right.\\[1.3ex]\left.\qquad\quad+(j-1/2)^2 \sqrt{1-kb^2 \left(j-1/2 \right)^2 }\right)+\frac{l(l+1) }{j^2},\\[1.8ex]
        \boldsymbol{C}_{i,i-1}=-\frac{( j+1/2)^2}{ j (j+1)}\times\\[1.3ex] \sqrt{[1-kb^2 \left(j+1/2\right)^2]\sqrt{(1-kb^2 j^2 )(1-kb^2 (j+1)^2)}},\\[1.8ex]
        \boldsymbol{C}_{NN}=\frac{(N-1/2)^2}{ N^2} \sqrt{1-kb^2 \left(N-1/2\right)^2} \sqrt{1-kb^2  N^2}.
    \end{cases}\\[2ex]
    &\begin{cases}
        \boldsymbol{T}_{ii}=[\mu ba]^2,\\[1.8ex]
        \boldsymbol{T}_{NN}=0.
    \end{cases}\\
\end{aligned}
\end{equation}
Defining the vector $\boldsymbol{\pi}_{lm}$ analogously to $\boldsymbol{\phi}_{lm}$, we can write the Hamiltonian in vector form as
\begin{equation}\label{eq:discretized_hamiltonian_vector_form}
        2H=2\sum_{lm}H_{lm}=\sum_{lm}\left\{\frac{\boldsymbol{\pi}_{lm}^2}{a^3}+a\boldsymbol{\phi}_{lm}^T\tilde{\boldsymbol{C}}\boldsymbol{\phi}_{lm}\right\}.
\end{equation}
Here, we have implicitly defined $H_{lm}$. Since the time-dependent part of $\tilde{\boldsymbol{C}}$ is diagonal, we can diagonalize $\tilde{\boldsymbol{C}}$ by diagonalizing $\boldsymbol{C}$, which is always possible because $\boldsymbol{C}$ is real and symmetric. We then perform an orthonormal transformation, $\boldsymbol{\phi}_{lm}' = U \boldsymbol{\phi}_{lm}$, such that
\begin{equation}\label{eq:gammatilde_def}
    \tilde{\boldsymbol{\Gamma}}^2 = U \tilde{\boldsymbol{C}} U^T=\left( \Gamma_i^2 + [\mu ba(t)]^2 \right) \delta_{ij},
\end{equation}
where $\Gamma_i^2$ are the eigenvalues of $\boldsymbol{C}$. Since $\boldsymbol{C}$ is not Toeplitz in spherical coordinates , $\Gamma_i^2$ should be computed numerically\footnote{Specifically, in Cartesian coordinates, $\boldsymbol{C}$ is tri-diagonal with each diagonal being constant. On the other hand, in spherical coordinates, $\boldsymbol{C}$ remains tri-diagonal, but its diagonals are not constant. Constant diagonals allow for recursive formulas to obtain the determinant and, consequently, enable the calculation of eigenvalues analytically. However, if the diagonals are not constant, an analytical closed-form solution for computing the eigenvalues is not generally known.}. For simplicity, we omit the explicit $l$-dependence in $\Gamma$, unless required. Thus, we can rewrite $H_{lm}$ as
\begin{equation}\label{eq:H_lm_explicit}
    \begin{aligned}
        2H_{lm} &= \frac{(\boldsymbol{\pi}_{lm}')^2}{a^3} + a(\boldsymbol{\phi}_{lm}')^T \tilde{\boldsymbol{\Gamma}}^2 \boldsymbol{\phi}_{lm}' \\
        &= \sum_{j=1}^N \biggl( \frac{ (\pi'_{lmj})^2}{a^3} + a \bigl[\Gamma_j^2+(\mu ba(t))^2\bigr] (\phi'_{lmj})^2 \biggr).
    \end{aligned}
\end{equation}

\subsection{Ground state of the discretized field}\label{sec:quantization_of_the_discret_field}

The Hamiltonian displayed in Eq.~(\ref{eq:H_lm_explicit}) can be quantized via the substitutions
\begin{equation}
    \begin{aligned}
        \pi'_{lmj} &\rightarrow \langle \phi'_{lmj}|\hat{\pi}'_{lmj}|\Psi\rangle = -i\hbar\frac{\partial}{\partial\phi'_{lmj}}\langle \phi'_{lmj}|\Psi\rangle,\\[1mm]
        \phi'_{lmj} &\rightarrow \langle \phi'_{lmj}|\hat{\phi}'_{lmj}|\Psi\rangle = \phi'_{lmj}\langle \phi'_{lmj}|\Psi\rangle,
    \end{aligned}
\end{equation}
where $\ket{\Psi}$ denotes the discretized field state vector, $\hat{\phi}$ and $\hat{\pi}$ represent the canonical operators of the field, and $ \phi'_{lmj}$ denote the discretized degrees of freedom of the field. Applying these quantization rules yields
\begin{equation}\label{eq:tdhos}
    \hat{H}_{lm} = \sum_{j=1}^N \left\{ \frac{(\hat{\pi}'_{lmj})^2}{2M(t)} + \frac{1}{2}M(t)\omega_j^2(t)(\hat{\phi}'_{lmj})^2 \right\},
\end{equation}
which describes a system of uncoupled harmonic oscillators with time-dependent mass and frequency. In particular, we define
\begin{align*}
    M(t)=a^3(t),\quad\text{and}\quad
    \omega_j=\frac{1}{a}\sqrt{\Gamma_j^2 + [\mu b a(t)]^2}.
\end{align*}

To determine the ground state of this Hamiltonian, one needs to solve the Ermakov-like equation \cite{Leach1977,Lohe2009,Ermakov2008}
\begin{equation}\label{eq:ermak_like_0}
    \ddot{\rho}_j +\frac{\dot{M}}{M}\dot{\rho}_j+\omega_j^2\rho_j=\frac{1}{M^2\rho_j^3}.
\end{equation}
Once $\rho_j$ is determined, the ground state of $H_{lm}$ (restoring the explicit $l$-dependence) is given by
\begin{equation}\label{eq:H_lm_ground_state}
    \begin{aligned}
        &\psi_0^{lm}=\langle \boldsymbol{\phi}_{lm}'|0\rangle \\
        &= \prod_{j=1}^{N} c_{lmj} \, e^{i\alpha_0^{lj}(t)} \exp\left[-\frac{1}{2}\left(\frac{1}{\rho_{lj}^2} - \frac{i M \dot{\rho}_{lj}}{\rho_{lj}}\right)(\phi'_{lmj})^2\right],
    \end{aligned}
\end{equation}
where $c_{lmj}$ are the wavefunction normalization constants and $\alpha_n^{lj}(t)$ are phase factors ensuring that the solution satisfies the Schrödinger equation. They are given by
\begin{equation*}
    \alpha_n^{lj}(t) = -\left(n + \frac{1}{2}\right) \int_{0}^{t} \frac{dt'}{M(t') \rho_{lj}^2(t')},
\end{equation*}
where $n$ is the energy quantum number. Since the total Hamiltonian in Eq.~(\ref{eq:discretized_hamiltonian_vector_form}) is separable, the complete ground state is given by
\begin{equation}\label{eq:product_ground_state_general_form}
    \Psi_0=\prod_{l=0}^{\infty}\prod_{m=-l}^{l}\psi_0^{lm}.
\end{equation}
\subsection{Analytical ground state solutions}

We now specify the above-presented techniques to the case of spherical collapse.

Remarkably, a generic solution, namely a solution valid in any general background, is not analytical. However, under precise conditions, useful to understand the behavior of the solution close to given regions, it is possible to argue analytical solution.

The main purpose of this subsection is to provide an example, under the form of analytical solution, that characterizes the ground state structure of our field.

To do so, let us consider the Ermakov-like relation in Eq.  (\ref{eq:ermak_like_0}) that can be recast as
\begin{equation}\label{eq:ermakov_equation_FLRW}
    \ddot{W} + \left( \omega^2(t) - \frac{9}{4}H^2(t) - \frac{3}{2}\dot{H}(t) \right) W = \frac{1}{W^3},
\end{equation}
where we have introduced $ W = \rho e^{\frac{g}{2}} $ and $ g=\ln(M) $. Further, $H=\dot{a}/a$ is the Hubble function, and $ \omega $ is one of the eigenvalues of the coupling matrix. Both $ W $ and $ \omega $ depend on $ l $ and $ j $, but we omit these indices for simplicity.

Defining $ f^2(t) = \omega^2(t) - \frac{9}{4}H^2(t) - \frac{3}{2}\dot{H}(t) $, the general solution of Eq.~(\ref{eq:ermak_like_0}) can be then expressed as
\begin{equation}\label{eq:ermakov_solution1}
    C_1 \rho^2 = \frac{z^2}{M} \left[ 1 + \left( C_2 + C_1 \int \frac{\mathrm{d}t}{z^2} \right)^2 \right],
\end{equation}
where $ z $ is any nontrivial solution of the equation
\begin{equation}\label{eq:classical_tdhoe}
    \ddot{z} =-f^2_{lj}(t) z= -\left[ \frac{1}{a^2(t)} \left( \Gamma_{lj}^2 + (\mu a(t) b )^2 \right) - \frac{9}{4}H^2 - \frac{3}{2}\dot{H} \right] z,
\end{equation}
thus having explicit dependence on $l$. Eq.~(\ref{eq:classical_tdhoe}) admits analytical solutions only in special cases. Furthermore, evaluating the integral in Eq.~(\ref{eq:ermakov_solution1}) poses significant challenges.

\subsubsection{Ground state in the Einstein-de Sitter background}

One of the rare cases in which an analytical solution of the Ermakov-like equation (\ref{eq:ermakov_equation_FLRW}) can be found is for the Einstein-de Sitter (EdS) universe. Although this metric cannot model gravitational collapse, it can approximate it when the spatial curvature is small or at times near the collapse, as we have shown in the footnote of Sec. \ref{sec:OS}. Notably, since $ H(t) = 2/(3t) $ and $ a(t) = \left( t/t_0 \right)^{2/3} $, the expression $ 9H^2/4 + 3\dot{H}/2 $ appearing in Eq.~(\ref{eq:classical_tdhoe}) vanishes. Consequently, $ f_{lj} = \omega_{lj} $, and the frequencies $ f_{lj} $ simplify to
\begin{align}
    f_{lj} = \omega_{lj} = \frac{c}{\left( t/t_0 \right)^{2/3}} \sqrt{\Gamma_{lj}^2 + \left[ \mu \left( \frac{t}{t_0} \right)^{2/3} b \right]^2}.
\end{align}
To find the ground state, we need to solve
\begin{equation}
    \ddot{z} = -\frac{c^2}{\left( \frac{t}{t_0} \right)^{4/3}} \left[ \Gamma_{lj}^2 + \left( \mu \left( \frac{t}{t_0} \right)^{2/3} b \right)^2 \right] z,
\end{equation}

which in the case of massless fields gives
\begin{equation}
    \ddot{z} = -\left( \frac{p_{lj}}{3^3} \right)^{2/3} \frac{z}{t^{4/3}},
\end{equation}
with $ \left( p_{lj}/27 \right)^{2/3} = \Gamma_{lj}^2 t_0^{4/3} $. A particular solution is
\[
z(t) = \cos\left( \sqrt[3]{p_{lj} t} \right) + \sqrt[3]{p_{lj} t} \, \sin\left( \sqrt[3]{p_{lj} t} \right).
\]
Substituting into Eq.~(\ref{eq:ermakov_solution1}), we obtain the following solution

\begin{figure}[htbp]
    \centering
    \includegraphics[width=1\columnwidth,clip]{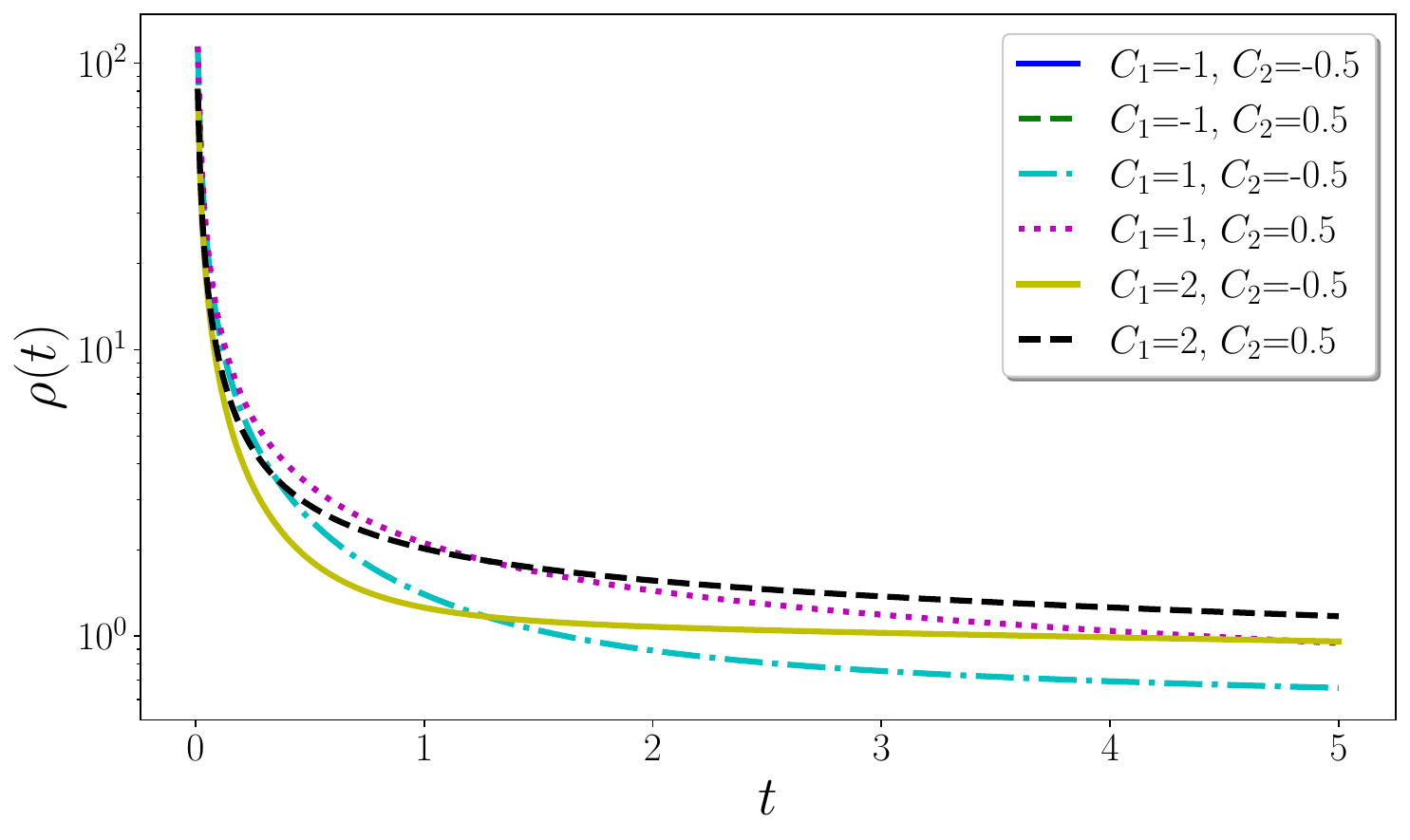}
    \includegraphics[width=1\columnwidth,clip]{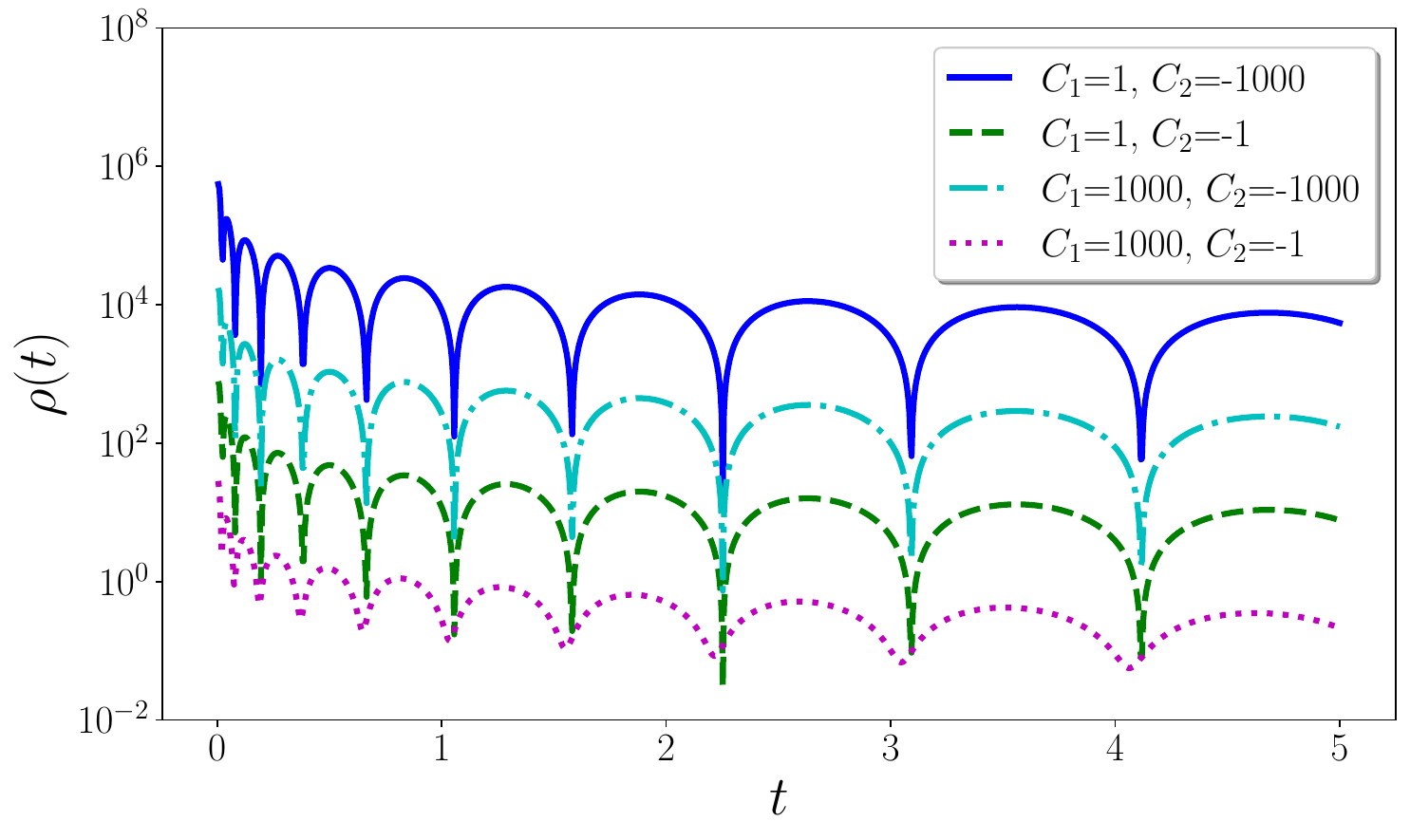}
    \caption{\textbf{Top:} Full solution for $\rho(t)$ in the EdS universe, with $p_{lj}=1$, $t_0=1$, and selected initial conditions. \textbf{Bottom:} Same as the top plot, but for high-frequency modes with $p_{lj}=10^4$.}
    \label{fig:exact_rho}
\end{figure}

\begin{widetext}
    \begin{equation} \label{eq:rhosolz}
    \begin{aligned}
         \rho &= \frac{|z| c}{\sqrt{C_1 a^3}} \left[ 1 + \left( C_2 + C_1 \int \frac{\mathrm{d}t}{z^2} \right)^2 \right]^{\frac{1}{2}} \\
         &= \frac{c \left| \cos\left( \sqrt[3]{p_{lj} t} \right) + \sqrt[3]{p_{lj} t} \, \sin\left( \sqrt[3]{p_{lj} t} \right) \right| }{\sqrt{C_1 \left( \frac{t}{t_0} \right)^2}}\left[ 1 + \left( C_2 + C_1 \left( -\frac{3 \left( p_{lj} \cos\left( \sqrt[3]{p_{lj} t} \right) \sqrt[3]{t} - p_{lj}^{2/3} \sin\left( \sqrt[3]{p_{lj} t} \right) \right) }{ p_{lj}^2 \sin\left( \sqrt[3]{p_{lj} t} \right) \sqrt[3]{t} + p_{lj}^{5/3} \cos\left( \sqrt[3]{p_{lj} t} \right) } \right) \right)^2 \right]^{\frac{1}{2}}.
    \end{aligned}
    \end{equation}
\end{widetext}
From Eq. \eqref{eq:rhosolz}, we notice that $ \rho $ cannot change sign during its evolution.
Fixing the initial conditions for $\rho$, namely $\rho(t_0; C_1,C_2)=\rho_0$ and $\dot{\rho}(t_0; C_1,C_2)=\dot{\rho}_0$, is analytically nontrivial. One could consider the low-frequency limit $p_{lj} \rightarrow 0$ to obtain an approximate solution for the low-frequency modes:
\begin{equation}
    z = 1 + \frac{1}{2} \left( p_{lj} t \right)^{2/3},
\end{equation}
which leads to the approximate expression
\begin{equation}\label{eq:approxrho}
    \begin{aligned}
         \rho &= \frac{1 + \frac{1}{2} \left( p_{lj} t \right)^{2/3}}{\sqrt{C_1 \left( \frac{t}{t_0} \right)^2}} \\
         &\quad \times \left\{ 1 + \left[ C_2 + C_1 t \left( 1 - \frac{3}{5} \left( p_{lj} t \right)^{2/3} \right) \right]^2 \right\}^{\frac{1}{2}}.
    \end{aligned}
\end{equation}
Similarly, choosing the initial conditions for the asymptotic solution of $\rho$ remains challenging. Therefore, we illustrate the solution by plotting it for specific values of $C_1$ and $C_2$. In Fig.~\ref{fig:exact_rho}, we display the full solution given by Eq.~(\ref{eq:rhosolz}). Note that the ground state of the discretized field, as given in Eq.~(\ref{eq:product_ground_state_general_form}), encompasses both low- and high-frequency modes: the low-frequency modes evolve via simple power laws, as one can deduce from Eq.(\ref{eq:approxrho}), whereas the high-frequency modes exhibit nontrivial oscillatory behavior.



\section{Entanglement entropy scaling of the ground state}\label{sec:ee_for_3Dgs}

In this section, we outline the procedure for computing the entanglement entropy scaling of the ground state, Eq.~(\ref{eq:product_ground_state_general_form}), for a time-dependent system of $N$ harmonic oscillators. We partition the oscillators into two sets:
\begin{itemize}
    \item \textbf{Inside Region ($ V $)}: Contains the $ n $ oscillators located within a specified closed volume $ V $, modeled as a three-dimensional sphere centered at the origin.
    \item \textbf{Outside Region}: Contains the remaining $ N - n $ oscillators outside this sphere.
\end{itemize}

Since the ground state in Eq.~(\ref{eq:product_ground_state_general_form}) is a pure state, the von Neumann entropy serves as an appropriate entanglement measure. By tracing out the degrees of freedom corresponding to the $ n $ inside oscillators, we obtain the reduced density matrix for the outside region. Notably, because the total system factorizes into radial and angular parts (labeled by $ l $ and $ m $), the partial trace over the inside region leaves the angular degrees of freedom factorized, while the radial degrees of freedom become mixed. In the density matrix representation, this can be expressed as
\begin{equation}\label{eq:product_state_after_trace}
   \rho_{\text{out}}^{(3\text{D})} = \text{Tr}_{V}[\rho^{(3\text{D})}] = \bigotimes_{l=0}^{\infty} \bigotimes_{m=-l}^{l} \rho^{(1\text{D})}_{\text{out}, lm},
\end{equation}
where $\rho^{(3\text{D})}=|\Psi_0\rangle \langle \Psi_0|$ is the density matrix representing the three-dimensional ground state, $\rho_{\text{out}}^{(3\text{D})}$ is the density matrix after tracing out the inside region, and $\rho^{(1\text{D})}_{\text{out}, lm}$ is the reduced density matrix for the radial mode associated with quantum numbers $ l $ and $ m $.

Since the von Neumann entropy is additive, namely $S(\rho_1 \otimes \rho_2) = S(\rho_1) + S(\rho_2)$, we focus on the one-dimensional Gaussian\footnote{This comes from the fact that $\Psi_0$ is Gaussian.} states $\rho^{(1\text{D})}_{\text{out}, lm}$. The total entropy is then given by
\begin{equation}\label{eq:eegs0}
    S(\rho_{\text{out}}^{(3\text{D})}) = \sum_{l=0}^{\infty} \sum_{m=-l}^{l} S\left(\rho^{(1\text{D})}_{\text{out}, lm}\right).
\end{equation}

Because the entanglement entropy of a Gaussian state depends only on the matrix defining its quadratic form, we denote this key object as the ground state matrix $\boldsymbol{\Sigma}^l$:
\begin{equation}\label{eq:ground_state_matrix}
    \Sigma_{\alpha\beta}^{l} = \left[\frac{1}{\rho_{l\alpha}^2} - iM\frac{\dot{\rho}_{l\alpha}}{\rho_{l\alpha}}\right]\delta_{\alpha\beta}.
\end{equation}
Thus, we have
\[
S\left(\rho^{(1\text{D})}_{\text{out}, lm}\right) = S\left(\boldsymbol{\Sigma}^l\right).
\]
Since $\boldsymbol{\Sigma}^l$ depends only on $l$, summing over $m$ yields
\begin{equation}\label{eq:ent_entropy_l_sum}
     S(\rho_{\text{out}}^{(3\text{D})}) = \sum_{l=0}^{\infty} (2l+1) S\left(\boldsymbol{\Sigma}^{l}\right).
\end{equation}

Although the computation of the entanglement entropy now resembles the time-independent scenario, the presence of an imaginary part in $\boldsymbol{\Sigma}^l$ introduces additional complications. Nonetheless, the full three-dimensional problem is reduced to a sum over independent one-dimensional problems indexed by $l$. The computation of the entanglement entropy for each one-dimensional time-dependent harmonic oscillator is presented below in Sec. \ref{appendix:ent_entro_compl_cov}.

The entanglement entropy depends on the number $ n $ of traced-out oscillators. Tracing out $ n $ oscillators corresponds to removing the first $ n $ spherical shells, with the radial coordinate given by $ r_n = bn $. To make this dependence explicit, we write
\begin{equation}\label{eq:ee_scaling}
    S(r_n) = \sum_{l=0}^{\infty} (2l + 1) S_n\left(\boldsymbol{\Sigma}^{l}\right),
\end{equation}
where $ S_n\left(\boldsymbol{\Sigma}^{l}\right) $ emphasizes the dependence on the radial coordinate. The total entropy then follows an area law if $ S(r_n) \propto r_n^2 $, a scaling behavior that has been demonstrated for the ground state of minimally coupled scalar fields in static spacetimes \cite{qft1,qft2}.

\subsection{Entanglement entropy of a complex ground state matrix}\label{appendix:ent_entro_compl_cov}
This section introduces a method for computing the scaling of entanglement entropy in the ground state of a time-dependent system. Specifically, our aim is to determine the entanglement entropy scaling for the state described in Eq.~(\ref{eq:product_ground_state_general_form}). Time dependence introduces an imaginary component within the ground state matrix. In the previous section \ref{sec:ee_for_3Dgs}, we demonstrated how to reduce the full three-dimensional case to an effective one-dimensional system. Here, we extend the procedure outlined in Appendix \ref{appendix:ee_real_cov_matrix}, to determine the scaling of entanglement entropy in a time-dependent, one-dimensional harmonic system.

Consider a Gaussian ground state in which the ground state matrix $\boldsymbol{\Sigma}^l$ is given by Eq. (\ref{eq:ground_state_matrix}), which we rewrite as
\begin{equation}
    \boldsymbol{\Sigma}(t) = \boldsymbol{\Sigma}_R(t) + i \boldsymbol{\Sigma}_I(t),
\end{equation}
where $ \boldsymbol{\Sigma}_R $ and $ \boldsymbol{\Sigma}_I $ are real matrices.
When employing normal coordinates, we know that $ \boldsymbol{\Sigma}(t) $ is diagonal. Let $ \boldsymbol{y} $ be the normal coordinate vector. Then the ground state is given by

\begin{equation}
    \Psi = \mathrm{Det}\left(\frac{\boldsymbol{\Sigma}_R}{\pi}\right)^{\frac{1}{4}} e^{-\frac{1}{2} \boldsymbol{y}^T \boldsymbol{\Sigma} \boldsymbol{y}},
\end{equation}
where $\Psi$ here plays the role of the state $\psi_0^{lm}$ defined in Sec. \ref{sec:ee_for_3Dgs}. The ground state density matrix is given by
\begin{equation}\notag
        \rho(\boldsymbol{y},\boldsymbol{y'})=\mathrm{Det}^{\frac{1}{2}}\left(\frac{\boldsymbol{\Sigma}_R}{\pi}\right) e^{-\frac{1}{2}\left( \boldsymbol{y}^T \boldsymbol{\Sigma} \boldsymbol{y}+\boldsymbol{y}'^T \boldsymbol{\Sigma}^* \boldsymbol{y}'\right)}.
\end{equation}
We can perform exactly the same steps as in Appendix \ref{appendix:ee_real_cov_matrix}, where the case of real ground state matrix is treated, taking onto account however also the complex conjugates. We arrive at a similar result to Eq.~(\ref{eq:useful_matrices_def}), which now reads
\begin{equation}
        \tilde{\boldsymbol{\gamma}}\coloneqq\boldsymbol{\gamma}+i \boldsymbol{\delta}\coloneqq\boldsymbol{D}-\frac{1}{2}\boldsymbol{B}^T\boldsymbol{A}_R^{-1}\boldsymbol{B},\qquad \boldsymbol{\beta}\coloneqq\frac{1}{2}\boldsymbol{B}^\dag\boldsymbol{A}_R^{-1}\boldsymbol{B},
\end{equation}
where $\boldsymbol{\beta}^\dag=\boldsymbol{\beta}$ and $\tilde{\boldsymbol{\gamma}}^T=\tilde{\boldsymbol{\gamma}}$.
With these definitions, we rewrite the density matrix as
\begin{equation}
    \begin{aligned}
        &\rho_{\text{out}}(\boldsymbol{\eta}_N,\boldsymbol{\eta}_N')=\mathrm{Det}^{\frac{1}{2}}\left(\pi \boldsymbol{A}_R^{-1}\right)\mathrm{Det}^{\frac{1}{2}}\left(\frac{\boldsymbol{\Sigma}_R}{\pi}\right)\times\\&\times e^{-\frac{1}{2}\left(\boldsymbol{\eta}_N^T\boldsymbol{\gamma}\boldsymbol{\eta}_N+\boldsymbol{\eta}_N'^T\boldsymbol{\gamma}\boldsymbol{\eta}_N'-2\boldsymbol{\eta}_N'^T\boldsymbol{\beta}\boldsymbol{\eta}_N\right)}e^{\frac{i}{2}\left(\boldsymbol{\eta}_N^T\boldsymbol{\delta}\boldsymbol{\eta}_N+\boldsymbol{\eta}_N'^T\boldsymbol{\delta}\boldsymbol{\eta}_N'\right)}.
    \end{aligned}
\end{equation}
The fact that $\tilde{\boldsymbol{\gamma}}$ is complex but not Hermitian does not allow to diagonalize it with an orthonormal basis change. However, according to \cite{Katsinis:2023hqn}, for the computation of the eigenvalues of the density matrix we can neglect its imaginary part: the matrix $\boldsymbol{\delta}$. So, setting $\boldsymbol{\delta}=0$ do not affect the eigenvalues and we can still proceed as in the flat spacetime scenario. We perform two consecutive coordinate transformations. The first transformation, $\boldsymbol{\eta} = \boldsymbol{V} \tilde{\boldsymbol{\eta}}$, diagonalizes the matrix $\boldsymbol{\gamma}$, resulting in $\boldsymbol{\gamma}_D = \boldsymbol{V}^T \boldsymbol{\gamma} \boldsymbol{V}$. The second transformation is defined by $\bar{\boldsymbol{\eta}} = \sqrt{\boldsymbol{\gamma}_D} \, \tilde{\boldsymbol{\eta}}$, which further simplifies the expression. After applying these orthogonal coordinate transformations, the out density matrix becomes
\begin{equation}\label{eq:out_density_matrix_simplified}
    \rho_{\text{out}}(\bar{\boldsymbol{\eta}}_N,\bar{\boldsymbol{\eta}}_N') \propto e^{-\frac{1}{2}\left(\bar{\boldsymbol{\eta}}_N^T \bar{\boldsymbol{\eta}}_N + (\bar{\boldsymbol{\eta}}_N')^T\bar{\boldsymbol{\eta}}_N' - 2\bar{\boldsymbol{\eta}}_N^T \tilde{\boldsymbol{\beta}} \bar{\boldsymbol{\eta}}_N'\right)},
\end{equation}
where
\begin{equation}
    \tilde{\boldsymbol{\beta}}=\boldsymbol{\gamma}_D^{-1/2} \boldsymbol{V}^T \boldsymbol{\beta} \boldsymbol{V} \boldsymbol{\gamma}_D^{-1/2}.
\end{equation}
In the flat spacetime case, we can perform an additional real orthonormal basis change that diagonalizes $\tilde{\boldsymbol{\beta}}$ \cite{Boutivas:2023ksg}. After this transformation, the reduced density matrix assumes a factorized form, making it straightforward to compute its eigenvalues. As one can see, the out density matrix only depends on the angular momentum index $l$, as the dependence is hidden in the definition of $\tilde{\boldsymbol{\beta}}$. Eq.(\ref{eq:out_density_matrix_simplified}) corresponds to $\rho_{lm,\text{out}}^{(1\text{D})}$ in Sec. \ref{sec:ee_for_3Dgs}, Eq.(\ref{eq:eegs0}).

However, in the current scenario, we cannot proceed as in flat spacetime because $\tilde{\boldsymbol{\beta}}$ is not real and symmetric but Hermitian. Although $\tilde{\boldsymbol{\beta}}$ is still diagonalizable, the change of basis is unitary rather than orthogonal. As a result, the reduced density matrix cannot be expressed as a product state. Nevertheless, we can still compute the eigenvalues of the reduced density matrix Eq.~(\ref{eq:out_density_matrix_simplified}), which are given by the $\min(n, N-n)$ solutions to the equation
\begin{equation}\label{eq:eigenvalues_red_dens_matrix}
\text{Det}\left(2\boldsymbol{I} - \lambda \tilde{\boldsymbol{\beta}} - \frac{1}{\lambda} \tilde{\boldsymbol{\beta}}^T\right) = 0.
\end{equation}
Let us denote the solutions of this equation by $\xi_i$. Due to the similarity of the eigenvalue structure to the time-independent case, the entanglement entropy can be expressed in the same manner as in flat spacetime
\begin{equation}\label{eq:entropy_scaling_time_dep}
    S_n(\rho_{\text{out}}) = -\sum_{i=1}^{N-n} \left\{ \ln(1 - \xi_i) + \frac{\xi_i}{1 - \xi_i} \ln \xi_i \right\},
\end{equation}
where $S_n(\rho_{\text{out}})$ is the von Neumann entropy of the out density matrix. This result has been used in Eq.(\ref{eq:ee_scaling}). In particular, we include the subscript $n$ to emphasize the dependence on the size of the inaccessible region. The $\xi_i$ are now functions of time, and thus the entropy depends on time as well.


\subsubsection{Finding the eigenvalues}

As discussed above, we need to solve the determinant equation, Eq.~(\ref{eq:eigenvalues_red_dens_matrix}), in order to find the entanglement entropy of the ground state. Accordingly, we first reformulate the problem into a quadratic eigenvalue problem, and then we linearize it into a generalized eigenvalue problem.

Instead of solving the determinant equation (\ref{eq:eigenvalues_red_dens_matrix}), we can equivalently solve the eigenvalue problem
\begin{equation}
    \left(2\boldsymbol{I} - \lambda \tilde{\boldsymbol{\beta}} - \frac{1}{\lambda} \tilde{\boldsymbol{\beta}}^T\right)\boldsymbol{v}=0.
\end{equation}
where $\boldsymbol{v}$ is an eigenvector. Multiplying both sides of the determinant equation by $\lambda$, we obtain the quadratic eigenvalue problem
\begin{equation}
    \left( -\lambda^2 \tilde{\boldsymbol{\beta}} + 2 \lambda \boldsymbol{I} - \tilde{\boldsymbol{\beta}}^T \right) \boldsymbol{v} = 0.
\end{equation}
The above equation is equivalent to
\begin{equation}
    \left[\lambda^2 \boldsymbol{\beta} - \lambda (2\boldsymbol{\gamma} )+ \boldsymbol{\beta}^T \right] \boldsymbol{z} = 0,
\end{equation}
where $\boldsymbol{z}$ is a different eigenvector but $\lambda$ is the same eigenvalue. To solve this quadratic eigenvalue problem, we linearize it by introducing a new vector $\boldsymbol{w}$ defined by  \[\boldsymbol{w}^T=(\lambda \boldsymbol{z},\boldsymbol{z}).\] Then, the quadratic eigenvalue problem is equivalent to the generalized eigenvalue problem
\begin{equation}\label{eq:generalized_eigenv_problem}
    \boldsymbol{F}\boldsymbol{w}=\lambda \boldsymbol{G} \, \boldsymbol{w},
\end{equation}
where
\[
\boldsymbol{F}=\begin{pmatrix}
        -2\boldsymbol{\gamma}&\boldsymbol{\beta}^T\\
        \boldsymbol{I}&\boldsymbol{0}
    \end{pmatrix},\qquad
    \boldsymbol{G}=\begin{pmatrix}
        -\boldsymbol{\beta}&\boldsymbol{0}\\
        0&\boldsymbol{I}\end{pmatrix}.
\]
The system Eq.~(\ref{eq:generalized_eigenv_problem}) can now be solved using numerical techniques such as the \texttt{eig} function from the \texttt{scipy} library, which computes the eigenvalues $\lambda$. The eigenvalues $\lambda$ obtained from solving Eq.~(\ref{eq:generalized_eigenv_problem}) correspond to the solutions of the original determinant equation (\ref{eq:eigenvalues_red_dens_matrix}).

\subsection{Summary on how to compute the entanglement entropy scaling}
To summarize, the procedure for computing the entanglement entropy scaling is as follows:
\begin{itemize}
    \item \textbf{Solve the Ermakov-like Equation:} For each eigenvalue $\Gamma_{lj}^2$ of $\boldsymbol{C}$,   (defined by Eqs.~(\ref{eq:coupling_matrix}) and (\ref{eq:gammatilde_def})), solve
    \begin{equation}\label{eq:init_cond}
        \begin{cases}
             \ddot{\rho}_{lj} + 3H\dot{\rho}_{lj} + \frac{1}{a^2(t)}\left[\Gamma_{lj}^2 + (\mu a(t) b)^2\right]\rho_{lj} = \frac{1}{a^6(t)\rho_{lj}^3}, \\
             \rho_{lj}(t_{0}) = \frac{c}{\left(a^3(t_{0})\sqrt{\Gamma_{lj}^2 + (\mu a(t_{0}) b)^2}\right)^{1/2}}, \\
             \dot{\rho}_{lj}(t_{0}) = 0,
        \end{cases}
    \end{equation}
    where $H=\dot{a}/a$ is the Hubble function and the initial conditions are chosen so that at $t_{0}$ the ground state coincides with the flat-spacetime ground state.

    \item \textbf{Define the Ground State Matrix:} Set
    \[
    \Sigma_{ij}^l(t) = \left[\frac{1}{\rho_{lj}^2} - iM\frac{\dot{\rho}_{lj}}{\rho_{lj}}\right]\delta_{ij}.
    \]

    \item \textbf{Compute the Entanglement Entropy:} Evaluate $S\left(\boldsymbol{\Sigma}^l(t)\right)$ following the method described in Section \ref{appendix:ent_entro_compl_cov}, and then use Eq.~(\ref{eq:ee_scaling}) to obtain the total entropy.
\end{itemize}

\section{Entanglement entropy during gravitational collapse}\label{sec:results}

We now aim to investigate the time evolution and the scaling of the ground state entanglement entropy for a scalar field propagating in the collapsing background outlined in Eq.~(\ref{eq:global_metric_comov}). In the attempt to interpret the entanglement entropy as the thermodynamic entropy of a black hole, we focus on the field degrees of freedom within the spatial region inside the collapsing sphere, $ r < r_b $. In this region, the background evolution is described by the spatially curved FLRW metric presented in Eq.~(\ref{eq:close_flrw_metric}). We then recall the action presented in Eq.~(\ref{eq:action_sf_flrw}), which describes the propagation of the scalar field within the collapsing region. We aim to compute the entropy measured by an observer which is at rest with respect to the free-falling matter. Since the global metric is given by Eq.~(\ref{eq:global_metric_comov}), we underline that a Novikov outside observer in free fall would measure the same entropy\footnote{We highlight that a different result would be obtained by a Schwarzschild observer, i.e., an observer at rest with respect to the center of the collapsing matter, located at spatial infinity. Below, we also provide a qualitative explanation of this scenario.}.

Table \ref{tab:collapse_parameters} presents the collapse parameters employed in our analysis, which are chosen to be compatible with numerical methods\footnote{All quantities are of order unity to prevent numerical algorithms from failing due to excessively large or small values.} and to ensure a non-negligible spatial curvature $ k $. If the collapse has not yet occurred then $ r_b \geq r_s $, which implies $ k \leq 1/r_s^2 $. Using a black hole mass of $3 M_\odot$, the lowest mass expected for an astrophysical black hole, we obtain the upper bound $ k \lessapprox 10^{-8}\, \text{m}^{-2} \approx 10^{-40}\, \text{GeV}^2 $. Thus, we expect low values for the scalar curvature in astrophysical black holes. On the other hand, primordial black holes could be, in principle, considerably smaller than astrophysical ones, so they may fall within the above-discussed mass range.

\begin{table}[ht]
    \centering
    \renewcommand{\arraystretch}{1.2}%
    \caption{OS Collapse Parameters}
    \label{tab:collapse_parameters}
    \begin{tabular}{|l|l|}
        \hline
        \hline
        \textbf{Parameter} & \textbf{Value} \\ \hline\hline
        $r_s\quad (\text{GeV}^{-1})$ & 1  \\ \hline
        $r_b\quad (\text{GeV}^{-1})$ & 2.0 \\ \hline
        $k \quad (\text{GeV}^{2})$ & 0.125 \\ \hline
        $t_{c}\quad (\text{GeV}^{-1})$ & 4.443 \\ \hline
        $t_{r_s}\quad (\text{GeV}^{-1})$ & 3.636 \\ \hline\hline
    \end{tabular}
\end{table}

Regarding the other parameters, we use $ b = r_b/N $, where $ N $ is the total number of spherical shells, and set $ \mu = 0 $ unless otherwise specified.

\begin{figure}[htbp]
    \centering
    \includegraphics[width=1\columnwidth,clip]{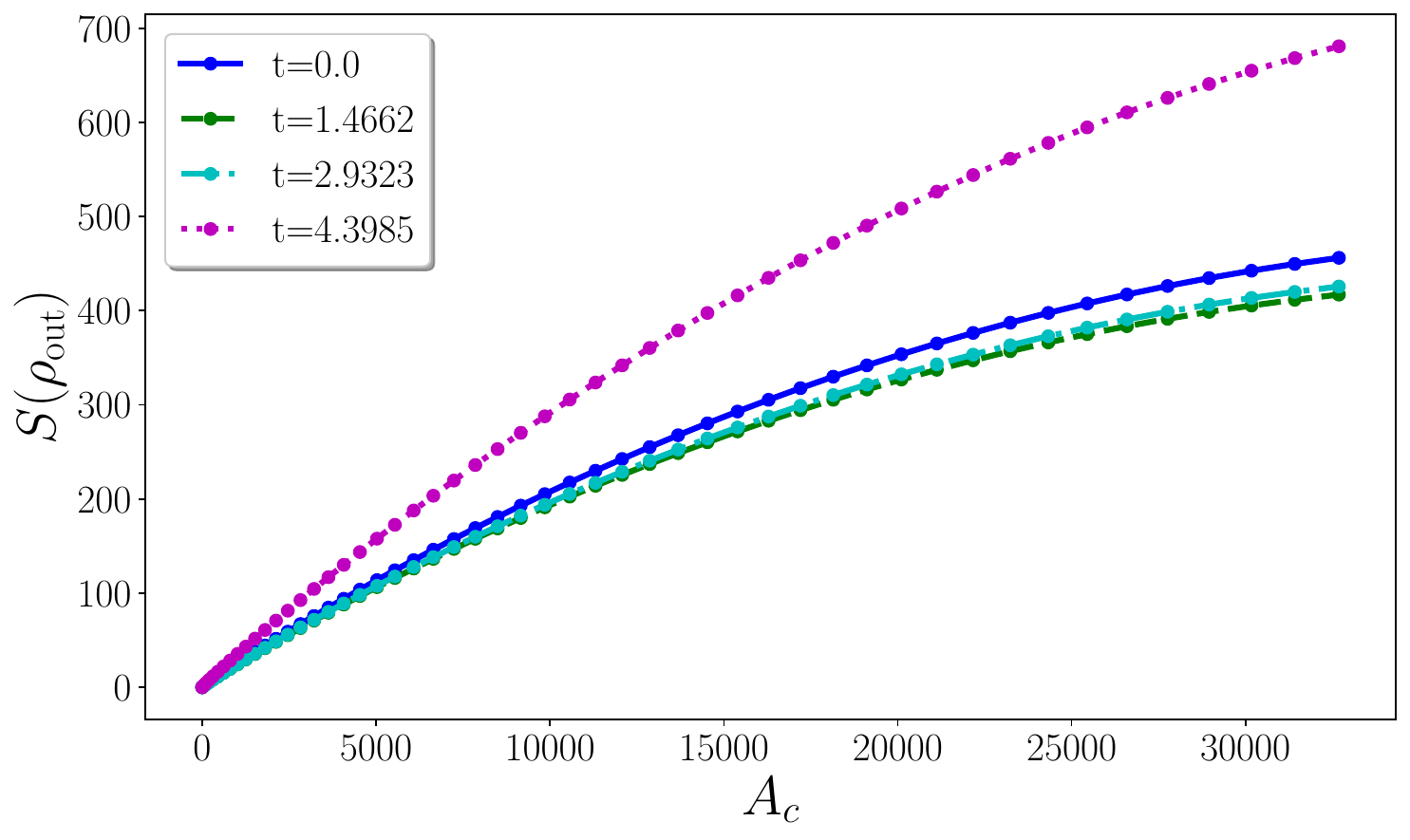}
    \includegraphics[width=1\columnwidth,clip]{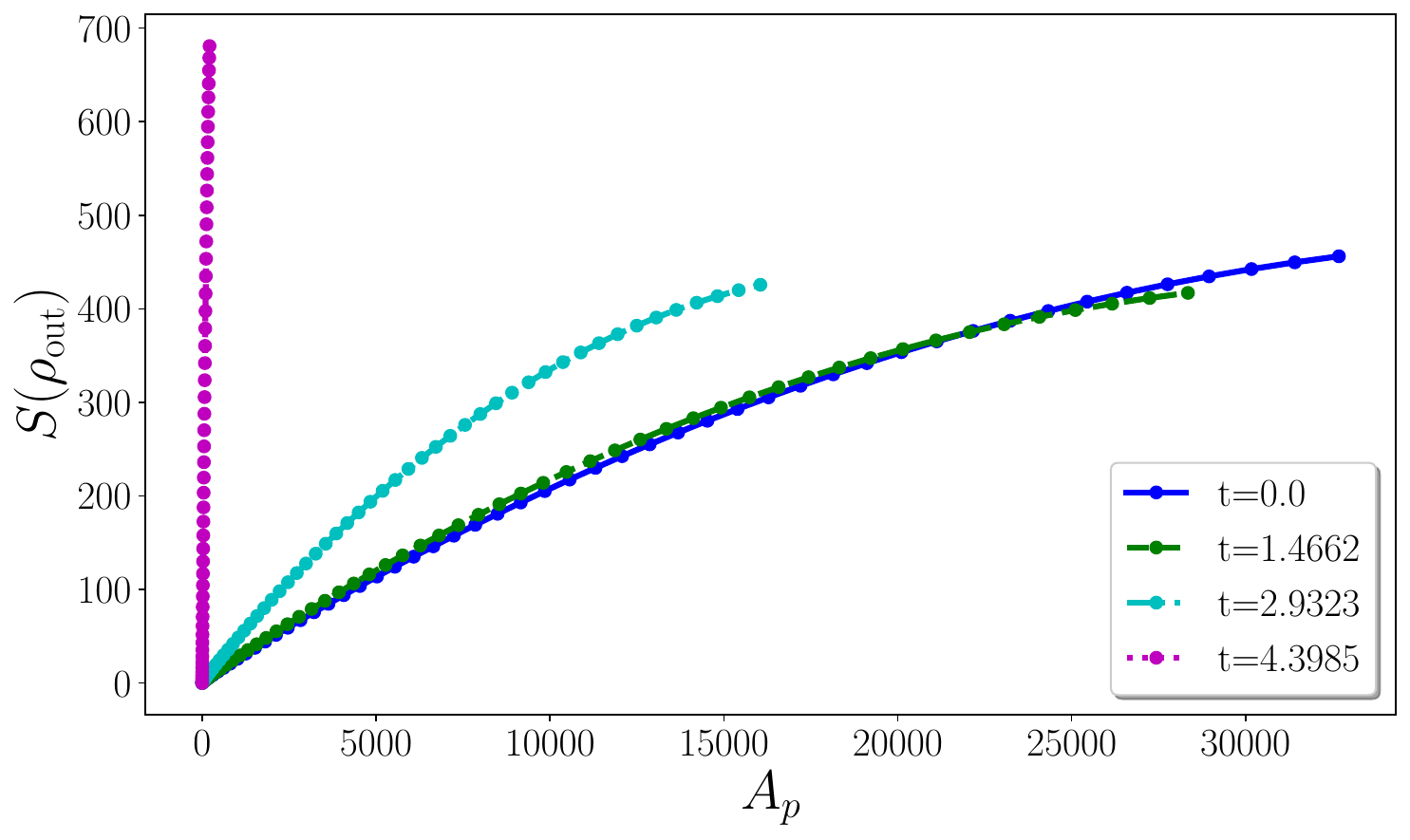}
    \caption{\textbf{Top}: Ground state entanglement entropy as function of the dimensionless comoving area $ A_c $ for various comoving times. The collapse parameters are reported in Table~\ref{tab:collapse_parameters}. We computed the entropy up to the 51\textsuperscript{st} shell ($n=51$) out of a total of 60 shells ($N=60$). We fixed $ l_{\mathrm{max}} = 1500 $ to achieve a tolerance of approximately $ 10^{-4}\% $, as discussed in \cite{Belfiglio:2024qsa}.
    \textbf{Bottom}: Ground state entanglement entropy as function of the physical area $ A_p = a^2 A_c $.  The parameters used are the same as in the top panel. The curves stop at a definite value of $A_p$ because the physical area is shrinking with the scale factor.  }
    \label{fig:eet0}
\end{figure}

We always consider the dimensionless area, defined as $ A_c = 4\pi n^2 $, where $ n $ is a positive integer and the subscript $c$ stands for comoving. The correct physical dimensions can be recovered by multiplying the dimensionless area by the squared cutoff parameter. We underline that all the above calculations are presented in comoving coordinates. Accordingly, the lengths and areas that an observer would measure are obtained by multiplying comoving lengths by the scale factor. Consequently, we define the dimensionless physical area as $ A_p = a^2(t) A_c $. The general relations between physical and comoving quantities are
\begin{equation}\label{eq:phy_com_relations}
    S(A_p) = S(a^2 A_c) \quad \text{and} \quad \frac{\partial S}{\partial A_p} = \frac{1}{a^2} \frac{\partial S}{\partial A_c},
\end{equation}
where $S$ is the entanglement entropy. We begin by verifying whether the area law holds within the OS collapsing background. From the numerical results presented in the top panel of Fig.~\ref{fig:eet0}, it is evident that the ground state entanglement entropy is not expected to follow an area law. The magnitude of the deviation can be traced back to the spatial curvature $ k $. Accordingly, when dealing with realistic astrophysical collapse scenarios, such deviations are typically less pronounced.
We evaluated the entanglement entropy at four distinct time points, selecting a sufficiently large number of shells. Our outcomes demonstrate that the entanglement entropy does not scale with the area of the region which is traced out, at each fixed comoving time. Nevertheless, sufficiently close to the origin, an area law is recovered, in agreement with a spatially flat EdS universe. This is guaranteed by the fact that Eq.~(\ref{eq:coupling_matrix}) reduces to the spatially flat EdS universe if $ k b^2 j^2 \ll 1 $. This suggests that the leading correction to the area law in the collapsing background, and possibly also in the Schwarzschild region, may be directly related to the nonzero spatial curvature found in the interior region. Referring to the top panel of Fig.~\ref{fig:eet0}, we observe that the entanglement entropy is not monotonic in time, as it initially decreases and then increases as the collapse progresses. Since this entropy is expressed in terms of the comoving area, an observer would not measure it directly. The same entropy is then shown in the bottom panel of Fig. \ref{fig:eet0} as function of the physical area $A_p$ for the same fixed times.

\begin{figure}[htbp]
    \centering
    \includegraphics[width=1\columnwidth,clip]{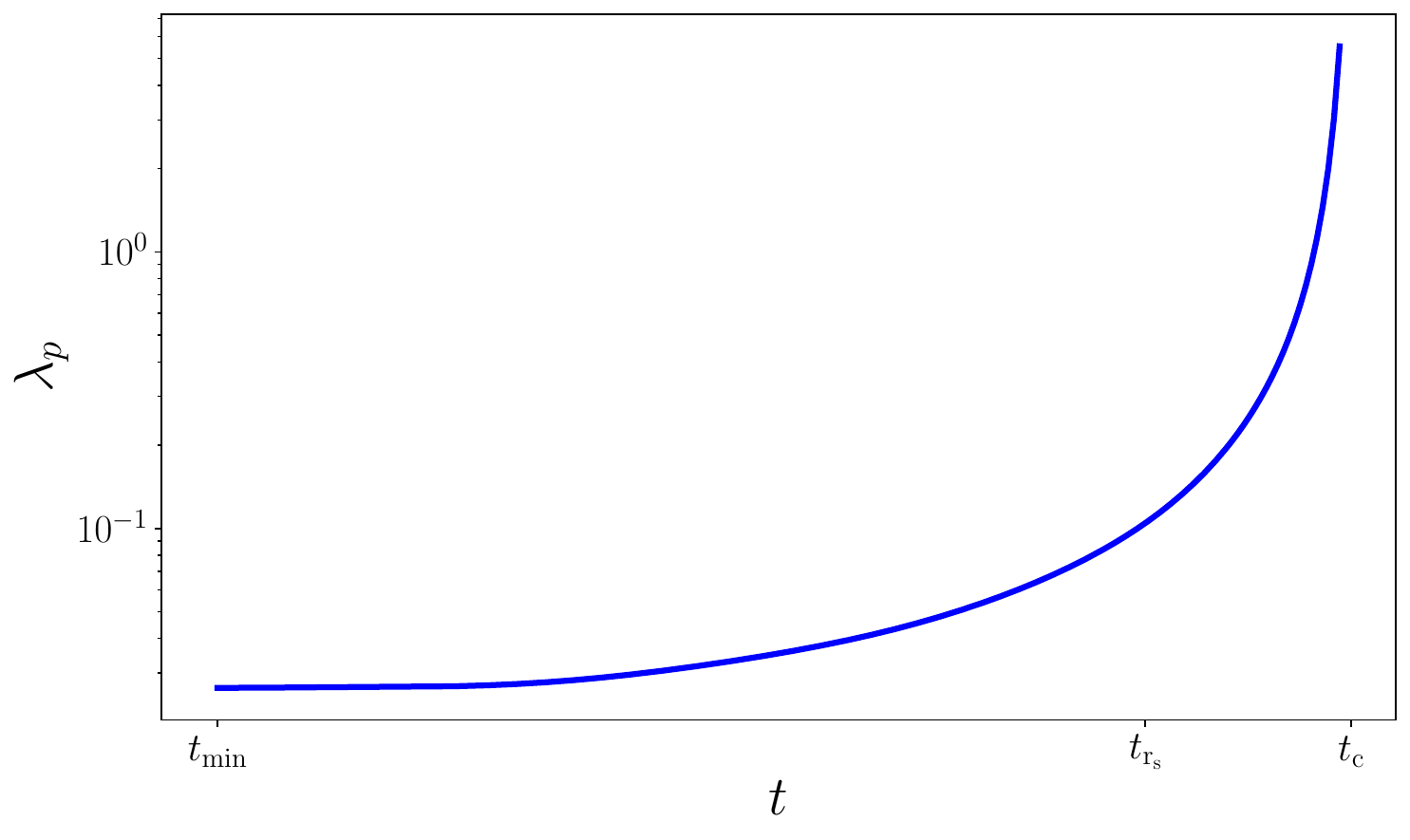}
    \includegraphics[width=1\columnwidth,clip]{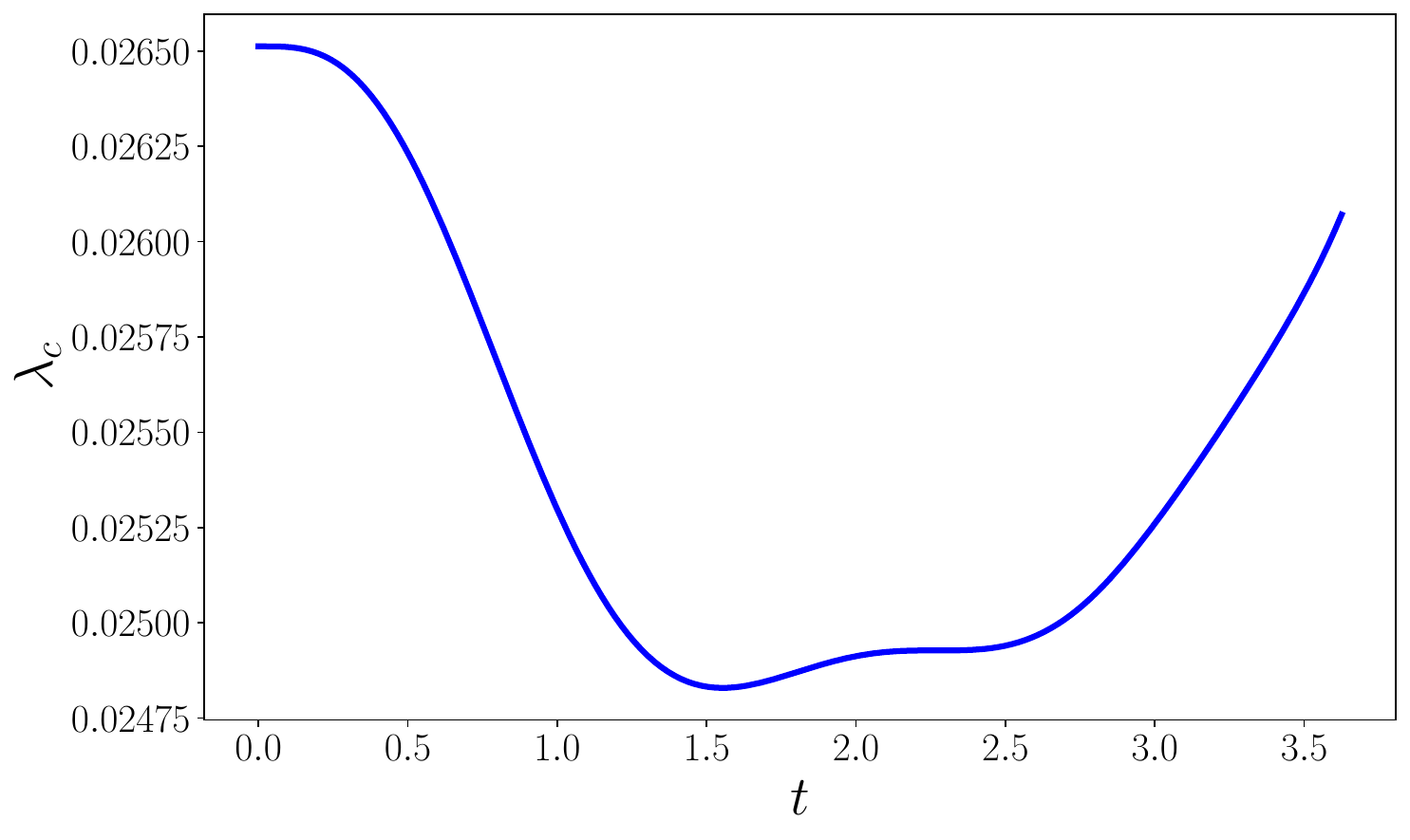}
    \caption{\textbf{Top}: Physical slope of the approximate area law near the origin, as a function of comoving time. The collapse starts at $ t = 0 $, and the sphere reaches the Schwarzschild radius at $ t_{r_s} = 3.636 $, while the singularity is reached at $ t_c = 4.443 $. \textbf{Bottom}: Comoving slope of the approximate area law near the origin as a function of comoving time. We restricted the time domain to $ t < t_{r_s} $ to highlight the oscillations that occur during the collapse. For both plots, we computed the entropy up to the $25$\textsuperscript{th} shell out of a total of $30$ shells. We fixed $ l_{\mathrm{max}} = 500 $ to achieve a tolerance of less than $ 0.01\% $. }
    \label{fig:eet0_slopes}
\end{figure}

As previously mentioned, the quantities that an actual observer would measure are the physical ones. Inspecting Eq.~(\ref{eq:phy_com_relations}), one can deduce that the physical slope diverges\footnote{It is understood that the entropy follows an area law. The general condition for having the divergence if the entropy asymptotically behaves as a power law $ a^\alpha $ is $ \alpha < 2 $.} as the collapse time is approached. This behavior can be intuitively inferred from Fig.~\ref{fig:eet0} and it is more precisely confirmed in Fig.~\ref{fig:eet0_slopes}.

In the presence of an area law, the slope of the entanglement entropy would be the sole parameter necessary to determine the entropy at any given shell or distance from the origin. This is precisely the case in the region near the origin, where the condition $ k b^2 j^2 \ll 1 $ is satisfied\footnote{With the parameters of Fig.~\ref{fig:eet0}, we have $ k b^2 j^2 \approx 10^{-2} \ll 1 $ for the first few shells. Furthermore, we numerically verified that the area law holds near the origin by computing the entropy scaling in this region.}. In this region, we have
\begin{equation}
    S = \lambda_p(t) A_p(t) = \lambda_c(t) A_c,
\end{equation}
where $ \lambda_c = \partial S / \partial A_c $. We underline that this region is relevant because it describes an approximately flat EdS cosmology. Thus, time-reversing the entanglement entropy in this region also describes the evolution of the entropy for a fixed comoving volume during the matter-dominated phase of the universe's history\footnote{Remarkably, we impose the initial condition $ H_0 = 0 $, which cannot be realized at any finite time if $ k = 0 $. Indeed, we are assuming that the system will evolve into a system of standard harmonic oscillators in flat spacetime after infinite time.}.

\begin{figure}[htbp]
    \centering
    \includegraphics[width=1\columnwidth,clip]{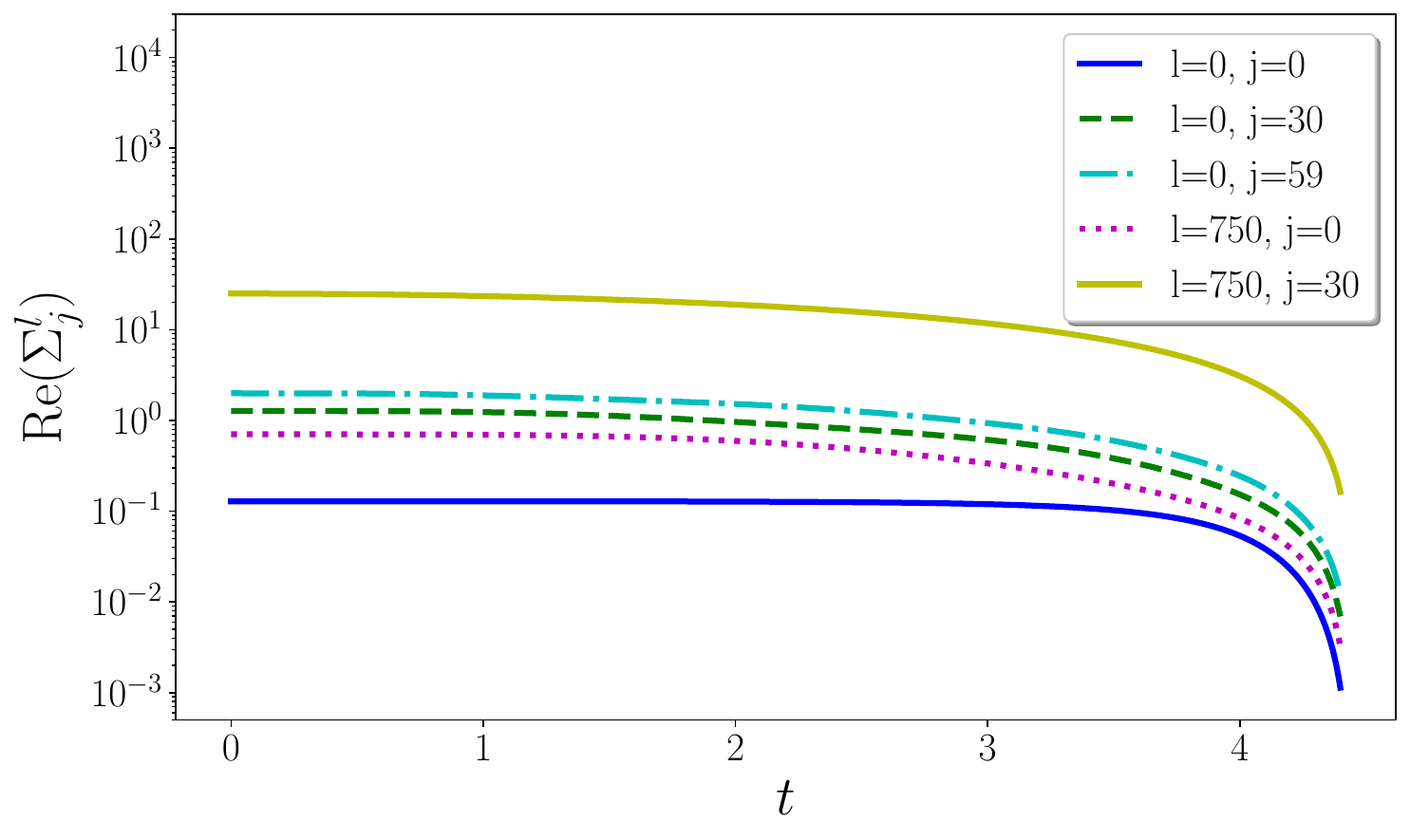}
    \includegraphics[width=1\columnwidth,clip]{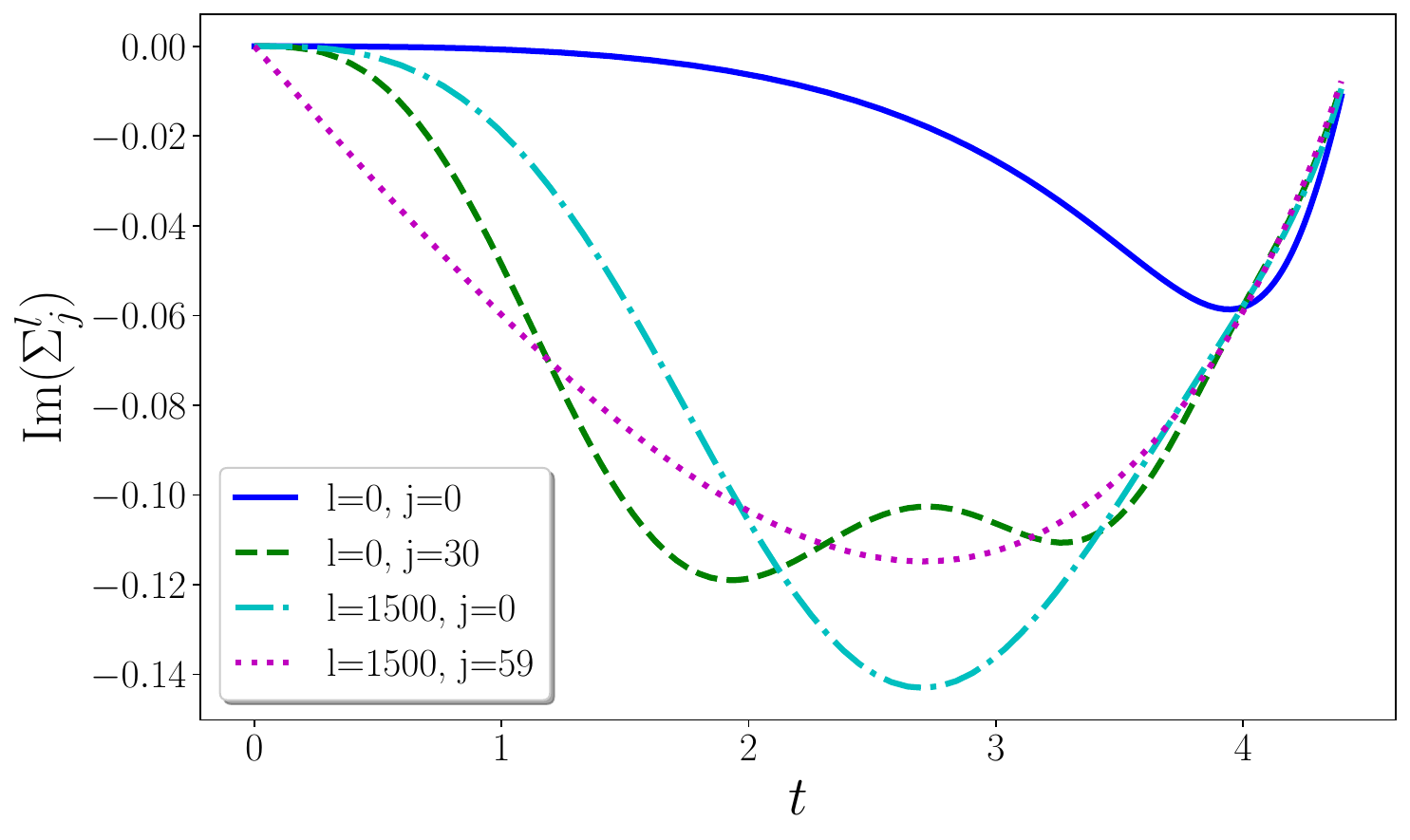}
    \caption{\textbf{Top}: Real part of the matrix elements $ \text{Re}(\Sigma_{j}^{l}) $. These are obtained by solving the Ermakov-like equation Eq.~(\ref{eq:ermak_like_0}) and using Eq.~(\ref{eq:ground_state_matrix}). They describe how the ground state of the field evolves in time for each angular momentum mode $ l $ and normal mode $ j $. \textbf{Bottom}: Imaginary part of the matrix elements $ \text{Im}(\Sigma_{j}^{l}) $.}
    \label{fig:ground_state_matrix}
\end{figure}

In Fig.~\ref{fig:eet0_slopes}, we present the time dependence of the slope of the entanglement entropy in the region where it approximately satisfies an area law during the collapse. The top plot displays the physical slope $ \lambda_p $ in the approximately spatially flat region. In the bottom plot, the comoving slope is shown. The observed oscillations can be attributed to the oscillatory character of the imaginary part of the ground state matrix, reported in the bottom plot of Fig.~\ref{fig:ground_state_matrix}.

A notable distinction in the entropy evolution is that, within the flat region, the entropy increases monotonically, whereas near the boundary, it can also decrease, as illustrated in the bottom plot of Fig.~\ref{fig:eet0}. Clearly, if the spatial curvature is very small, we would observe a monotonic entropy up to the boundary of the matter sphere.

\begin{figure}[htbp]
    \centering
    \includegraphics[width=1\columnwidth,clip]{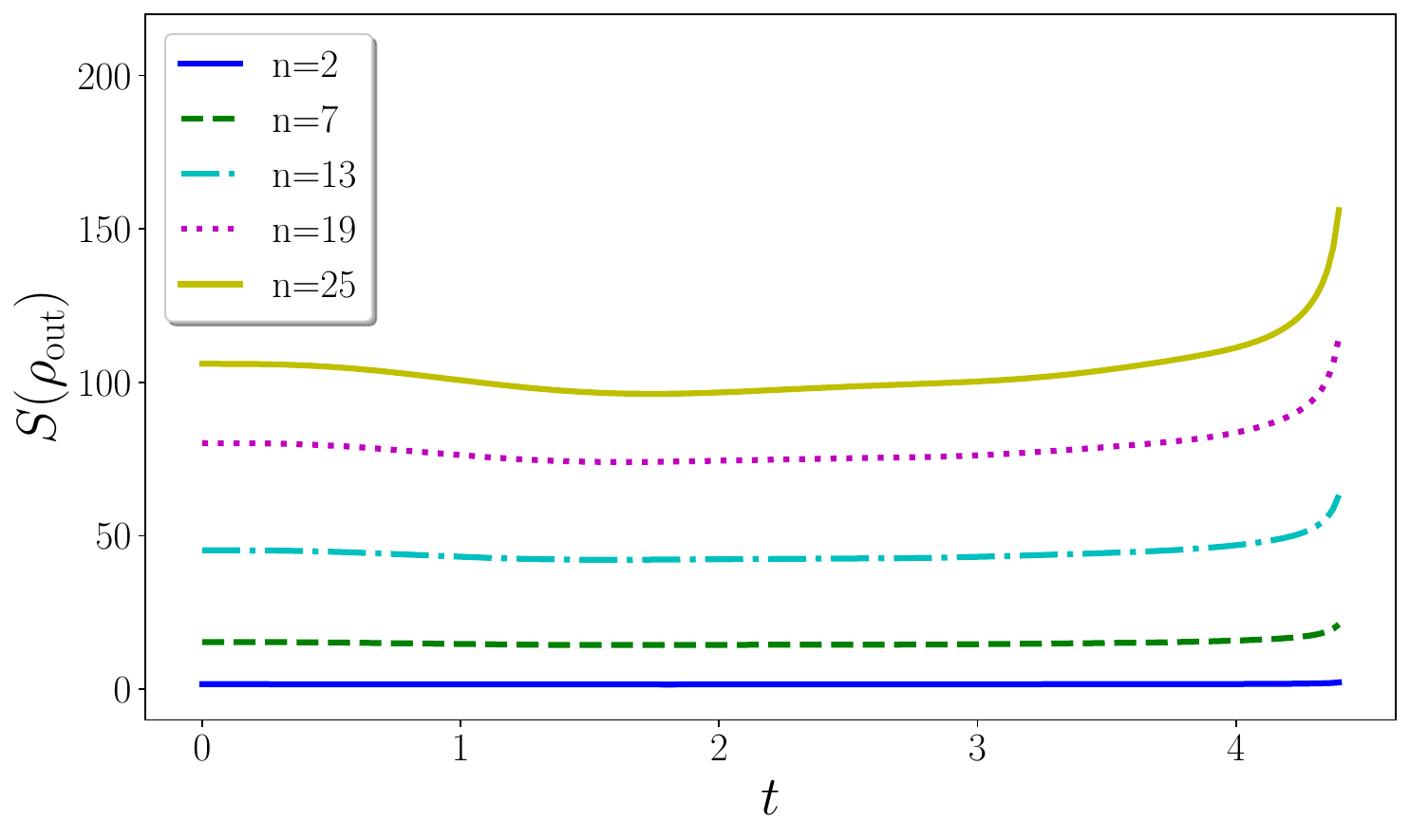}
    \includegraphics[width=1\columnwidth,clip]{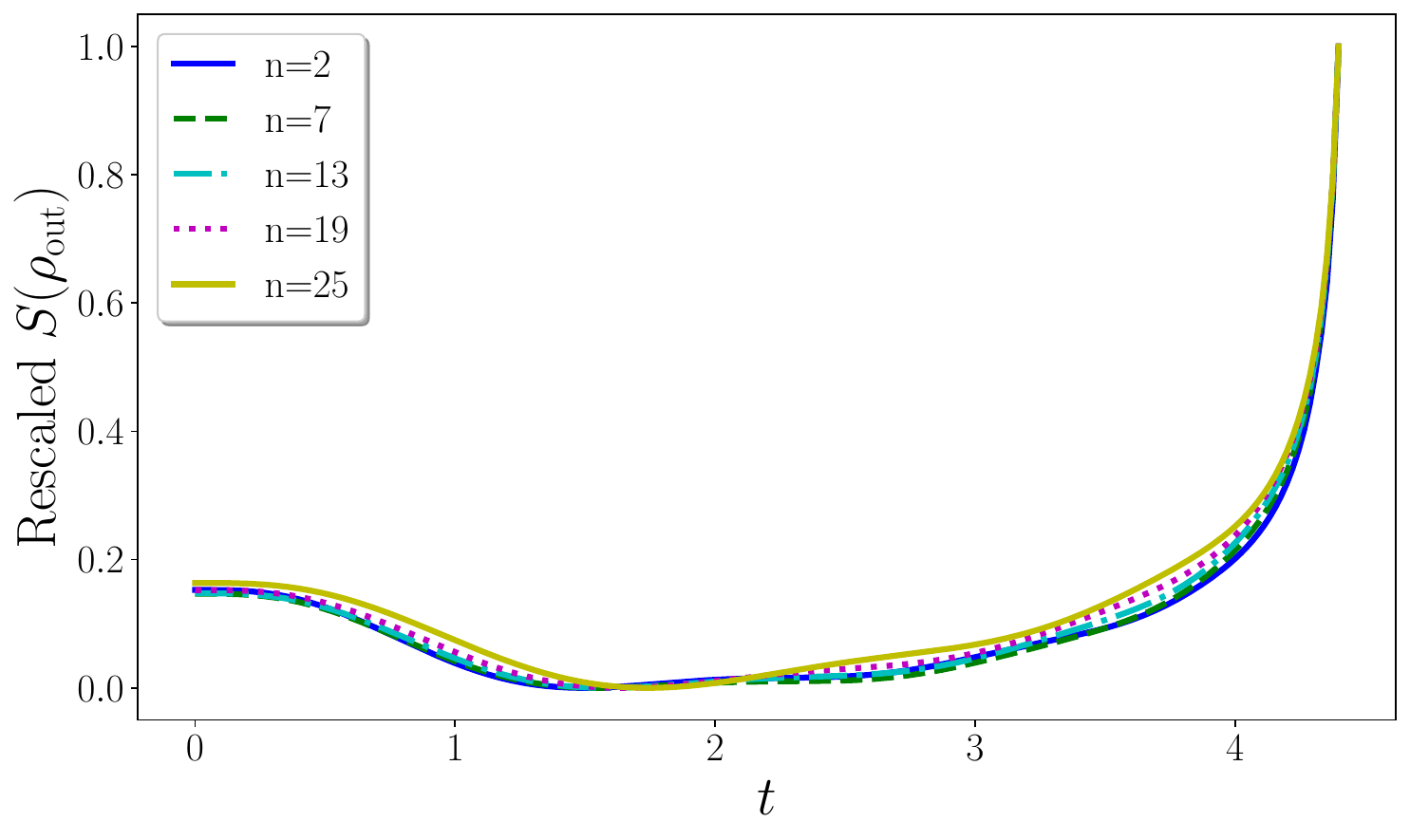}
    \caption{Ground state entanglement entropy computed at a given shell as a function of time. In the top plot, we present the real entropy values, while in the bottom plot, we show the rescaled and shifted curves to facilitate a fair comparison of the functional dependence of the entropy on time for the chosen shells. The rescaling ensures that each entropy reaches unity at the final time. The parameters are the same used in Fig. \ref{fig:eet0_slopes}.}
    \label{fig:ee_fixed_shell}
\end{figure}

The slope appears to diverge as $ t $ approaches $ t_c $. Importantly, a divergent slope does not necessarily imply that the entropy itself diverges in our context. However, Fig.~\ref{fig:ee_fixed_shell}, which depict the ground state entanglement entropy computed at a given shell as a function of time, suggests that the entropy may also diverge. From the bottom plot, we can also deduce that the shell we fix does not have an huge impact on the functional form of the entropy.
At $ t = t_c $, when all matter has collapsed into the singularity, the system ceases to exist as space contracts into a point. In this limit, the here-employed mathematical framework becomes inapplicable, and thus our numerical simulations are unable to accurately depict the final stages of collapse.

As usual, we expect that quantum gravity effects would intervene to prevent the formation of a singularity, thereby ensuring that the entanglement entropy remains continuous. For instance, Ref.~\cite{Malafarina:2022wmx} presents semiclassical collapse models that may lead to regular black hole solutions. Nonetheless, Figs.~\ref{fig:eet0}, ~\ref{fig:eet0_slopes} and ~\ref{fig:ee_fixed_shell} indicate that spacetime contraction during collapse inevitably affects entanglement entropy within the interior spatial region.

We also aim to understand the amount of entropy measured by an observer located far from the gravitational well. In the coordinate system we use, no event horizon is present. However, as demonstrated in Sec.~\ref{sec:OS}, the physical radius of the sphere, $ R(t) = a(t) r_b $, will ultimately decrease below the Schwarzschild radius, indicating the formation of a black hole. To determine what an external observer would measure in Schwarzschild coordinates, we move back the global metric in Schwarzschild coordinates, $ R, T $, defined in Eq.~(\ref{eq:schwarzschild_metric}), instead of the comoving coordinates, $ r, t $, employed thus far. Unfortunately, the interior metric in these coordinates does not allow an analytic computation of the entropy. For now, we defer solving the complete problem to future investigations. Instead of addressing the problem exactly, we compute the Schwarzschild time as a function of the comoving one, by integrating Eq.~(\ref{eq:dT_dt}). As mentioned in Sec.~\ref{sec:OS}, for an external Schwarzschild observer, the full collapse necessarily requires an infinite amount of time. Therefore, the time interval $ t > t_{r_s} $ is not observable by a Schwarzschild observer, as the collapsing sphere takes an infinite time to reach the Schwarzschild radius.

\begin{figure}[htbp]
    \centering
    \includegraphics[width=1\columnwidth,clip]{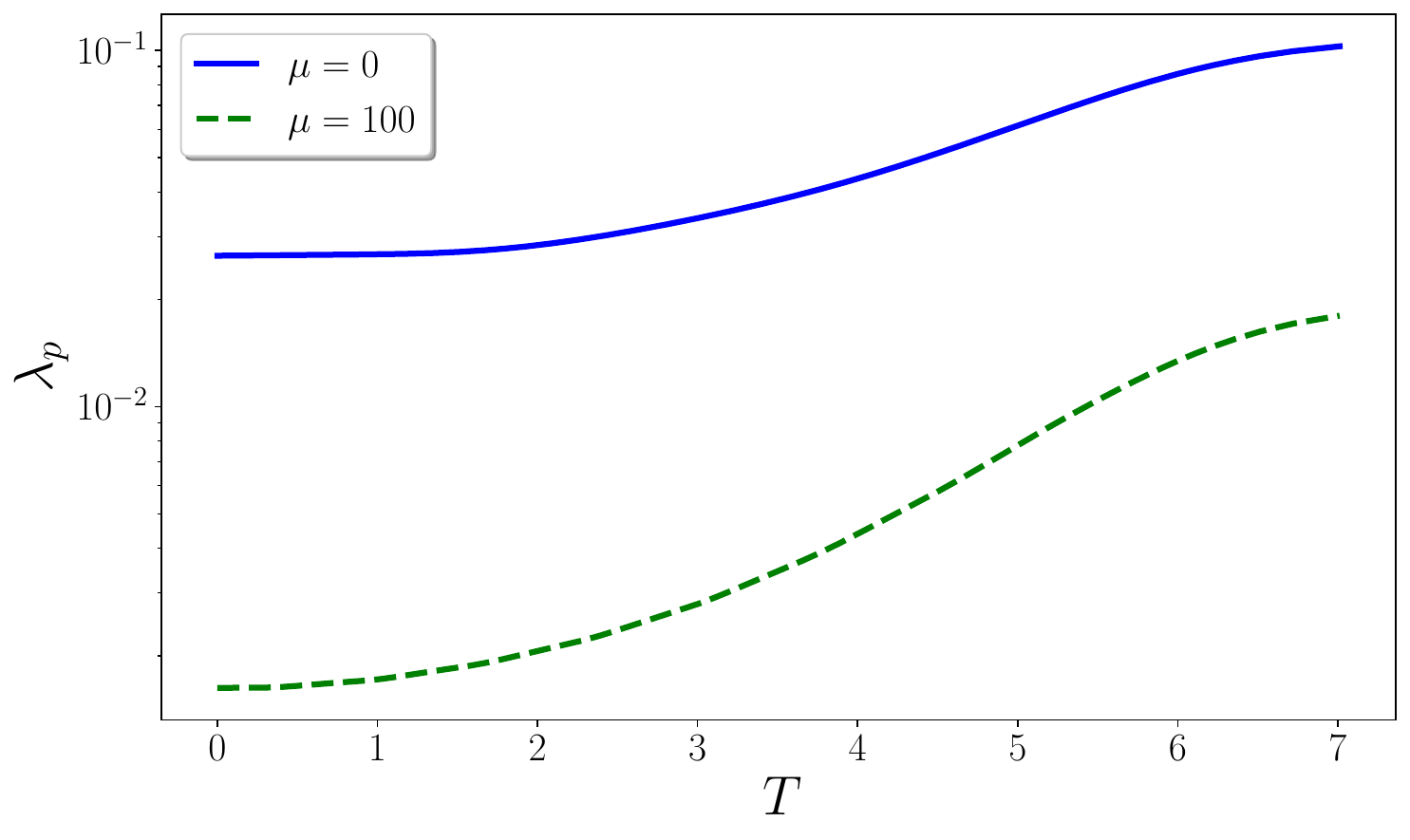}
    \caption{The entropy is identical to that in Fig.~\ref{fig:eet0_slopes}, but it is plotted against the time measured by an external observer using standard Schwarzschild coordinates. The entropy asymptotically approaches $ S(t_{r_s}) $.}
    \label{fig:eeT}
\end{figure}

Fig.~\ref{fig:eeT}  shows the entanglement entropy computed in comoving coordinates as function of the standard Schwarzschild time coordinate. The $\mu=0$ entropy is the same as the one shown in Fig.~\ref{fig:eet0_slopes}, while in the other we increased the field mass. It should be noted that Fig.~\ref{fig:eeT} provides a qualitative description of the entropy scaling, since a more refined approach would require the computation the entanglement entropy in Schwarzschild coordinates.

Nevertheless, the divergence of $ T(t) $ as $ t \rightarrow t_{\text{sch}} $ indicates that a standard Schwarzschild observer would perceive the entanglement entropy approaching an asymptotic value as the physical matter boundary radius approaches the Schwarzschild radius\footnote{This divergence can be deduced by inspecting Eq.~(\ref{eq:dT_dt}).}. Thus, despite Fig.~\ref{fig:eeT} lacks quantitative precision, it effectively captures the expected qualitative behavior of the entanglement entropy.

\section{Final remarks and outlooks}\label{sec:conclusions}

In this study, we investigated the dynamics of the ground state entanglement entropy for a discretized scalar field within the OS gravitational collapse model. Following standard approaches, the ground state has been derived as a product of Gaussian states associated with a set of uncoupled harmonic oscillators with time-dependent masses and frequencies. We initially solved the Ermakov-like equations to determine the ground state of the system, then computing the entanglement entropy associated with different spatial partitions during the collapse.

We observed that, within astrophysical collapse scenarios, the entanglement entropy exhibits several novel features with respect to static spacetimes. In particular, deviations from an area law typically arise in the presence of spatial curvature effects, while in the limit of a spatially flat spacetime ($k=0$) the area law is recovered, and only minor deviations are observed in realistic collapse scenarios characterized by negligible curvature.

Furthermore, we highlighted that the entanglement entropy exhibits a non-monotonic evolution in comoving time. During the early stages of collapse, the entropy decreases, while it becomes larger as the system approaches the central singularity. This oscillatory behavior can be traced back to the interplay between the real and imaginary components of the ground state matrix corresponding to the discretized field.

As the collapse proceeds toward the singularity, both the slope of the entanglement entropy with respect to the physical area and the entropy itself are expected to diverge. This divergence emphasizes the need for a quantum gravitational treatment to fully describe the physics of gravitational collapse and thus the corresponding entanglement features.

We also underlined that the perceived evolution of the entanglement entropy is observer-dependent. In particular, for an external Novikov (free-falling) observer, the amount of entanglement entropy is consistent with calculations in comoving coordinates. In contrast, a Schwarzschild observer at spatial infinity perceives the collapse as occurring over an infinite duration, such that the entanglement entropy asymptotically approaches a finite value. This discrepancy highlights the influence of the observer's frame of reference on the interpretation of the collapse dynamics.

Our study suggests that quantum entanglement can play an active role in addressing the physics of gravitational collapse, including possible connections to black hole thermodynamics and horizon formation, which have been widely studied in static scenarios. In particular, deviations from the area law and the increasing amount entanglement near singularities may help to illuminate how quantum correlations develop in strong-gravity regimes. The here-observed divergences hint at deeper physics, thus suggesting that a more comprehensive quantum gravity framework is likely required to fully understand the ultimate fate of field entanglement near singularities.

Looking ahead, several directions remain open: while our analysis treated the ground-state configuration in the interior region, an important extension involves applying appropriate boundary conditions for the quantum field at the horizon, in order to match them with the exterior Schwarzschild geometry. Such matching may alter or refine the entropy profile near the horizon as it forms. Furthermore, in order to address the divergence in the latest stages of collapse, it is necessary to move beyond the here-presented classical treatment of background geometry; semiclassical collapse models may offer better insight into near-singularity dynamics, and some quantum generalizations of the OS model have been proposed in recent years. At the same time, further studies might also clarify whether these entanglement profiles may offer insights into long-standing issues related to black hole information and holographic dualities.

Additionally, it would be interesting to provide an exact computation of the entanglement entropy for an external Schwarzschild observer, in order to determine whether the observed deviations from the area law mainly arise from using comoving coordinates or from the system’s intrinsic time dependence. More generally, our results show a complex entanglement structure in the presence of a dynamical background, thus representing a step toward a deeper understanding of quantum fields in strong-gravity regimes. By refining numerical and analytical tools to account for both spatial curvature and time-dependent processes, we expect that quantum information tools may play a key role in enhancing our understanding of black holes and quantum gravity.

\section*{Acknowledgements}

AB is thankful to Roberto Franzosi for support during the period in which this article has been written. OL acknowledges financial support from the  Fondazione  ICSC, Spoke 3 Astrophysics and Cosmos Observations. National Recovery and Resilience Plan (Piano Nazionale di Ripresa e Resilienza, PNRR) Project ID CN$\_$00000013 "Italian Research Center on  High-Performance Computing, Big Data and Quantum Computing"  funded by MUR Missione 4 Componente 2 Investimento 1.4: Potenziamento strutture di ricerca e creazione di "campioni nazionali di R$\&$S (M4C2-19 )" - Next Generation EU (NGEU)
GRAB-IT Project, PNRR Cascade Funding
Call, Spoke 3, INAF Italian National Institute for Astrophysics, Project code CN00000013, Project Code (CUP): C53C22000350006, cost center STI442016. S.M. and S.T. acknowledge financial support from "PNRR MUR project PE0000023-NQSTI".

\bibliographystyle{unsrt}

\begin{thebibliography}{10}

\bibitem{qit}
M.~M. Wilde.
\newblock {\em Quantum Information Theory}.
\newblock Cambridge University Press, Cambridge, 2017.

\bibitem{pl1}
M.~B. Plenio and S.~Virmani.
\newblock An introduction to entanglement measures.
\newblock {\em Quant. Inf. Comp.}, 7:1, 2007.

\bibitem{qft1}
L.~Bombelli, R.~K. Koul, J.~Lee, and R.~Sorkin.
\newblock Quantum source of entropy for black holes.
\newblock {\em Phys. Rev. D}, 34:373, 1986.

\bibitem{qft2}
M.~Srednicki.
\newblock Entropy and area.
\newblock {\em Phys. Rev. Lett}, 71:666, 1993.

\bibitem{rev1}
J.~Eisert, M.~Cramer, and M.~B. Plenio.
\newblock Area laws for the entanglement entropy - a review.
\newblock {\em Rev. Mod. Phys.}, 82:277, 2010.

\bibitem{PhysRevLett.93.227205}
F.~Verstraete, D.~Porras, and J.~I. Cirac.
\newblock Density matrix renormalization group and periodic boundary conditions: A quantum information perspective.
\newblock {\em Phys. Rev. Lett.}, 93:227205, Nov 2004.

\bibitem{PhysRevLett.94.060503}
M.~B. Plenio, J.~Eisert, J.~Drei\ss{}ig, and M.~Cramer.
\newblock Entropy, entanglement, and area: Analytical results for harmonic lattice systems.
\newblock {\em Phys. Rev. Lett.}, 94:060503, Feb 2005.

\bibitem{PhysRevB.73.085115}
M.~B. Hastings.
\newblock Solving gapped hamiltonians locally.
\newblock {\em Phys. Rev. B}, 73:085115, Feb 2006.

\bibitem{PhysRevLett.96.220601}
F.~Verstraete, M.~M. Wolf, D.~Perez-Garcia, and J.~I. Cirac.
\newblock Criticality, the area law, and the computational power of projected entangled pair states.
\newblock {\em Phys. Rev. Lett.}, 96:220601, Jun 2006.

\bibitem{RevModPhys.80.517}
Luigi Amico, Rosario Fazio, Andreas Osterloh, and Vlatko Vedral.
\newblock Entanglement in many-body systems.
\newblock {\em Rev. Mod. Phys.}, 80:517--576, May 2008.

\bibitem{bh1}
J.~M. Bardeen, B.~Carter, and S.~W. Hawking.
\newblock The four laws of black hole mechanics.
\newblock {\em Commun. Math. Phys.}, 31:161, 1973.

\bibitem{bh2}
J.~D. Bekenstein.
\newblock Black holes and entropy.
\newblock {\em Phys. Rev. D}, 7:2333, 1973.

\bibitem{bh3}
G.~'t~Hooft.
\newblock On the quantum structure of a black hole.
\newblock {\em Nucl. Phys.}, B256:727, 1985.

\bibitem{bh4}
T.~M. Fiola, J.~Preskill, A.~Strominger, and S.~P. Trivedi.
\newblock Black hole thermodynamics and information loss in two dimensions.
\newblock {\em Phys. Rev. D}, 50:3987, 1994.

\bibitem{Wald:1999vt}
Robert~M. Wald.
\newblock {The thermodynamics of black holes}.
\newblock {\em Living Rev. Rel.}, 4:6, 2001.

\bibitem{Page:2004xp}
Don~N. Page.
\newblock {Hawking radiation and black hole thermodynamics}.
\newblock {\em New J. Phys.}, 7:203, 2005.

\bibitem{PasqualeCalabrese_2004}
Pasquale Calabrese and John Cardy.
\newblock Entanglement entropy and quantum field theory.
\newblock {\em Journal of Statistical Mechanics: Theory and Experiment}, 2004(06):P06002, jun 2004.

\bibitem{Casini_2009}
H~Casini and M~Huerta.
\newblock Entanglement entropy in free quantum field theory.
\newblock {\em Journal of Physics A: Mathematical and Theoretical}, 42(50):504007, dec 2009.

\bibitem{RevModPhys.90.045003}
Edward Witten.
\newblock Aps medal for exceptional achievement in research: Invited article on entanglement properties of quantum field theory.
\newblock {\em Rev. Mod. Phys.}, 90:045003, Oct 2018.

\bibitem{holo}
R.~Bousso.
\newblock The holographic principle.
\newblock {\em Rev. Mod. Phys.}, 74:825, 2002.

\bibitem{PhysRevLett.96.181602}
Shinsei Ryu and Tadashi Takayanagi.
\newblock Holographic derivation of entanglement entropy from the anti--de sitter space/conformal field theory correspondence.
\newblock {\em Phys. Rev. Lett.}, 96:181602, May 2006.

\bibitem{Nishioka:2009un}
Tatsuma Nishioka, Shinsei Ryu, and Tadashi Takayanagi.
\newblock {Holographic Entanglement Entropy: An Overview}.
\newblock {\em J. Phys. A}, 42:504008, 2009.

\bibitem{Rangamani:2016dms}
Mukund Rangamani and Tadashi Takayanagi.
\newblock {\em {Holographic Entanglement Entropy}}, volume 931.
\newblock Springer, 2017.

\bibitem{RevModPhys.90.035007}
Tatsuma Nishioka.
\newblock Entanglement entropy: Holography and renormalization group.
\newblock {\em Rev. Mod. Phys.}, 90:035007, Sep 2018.

\bibitem{PhysRevD.73.121701}
Saurya Das and S.~Shankaranarayanan.
\newblock How robust is the entanglement entropy-area relation?
\newblock {\em Phys. Rev. D}, 73:121701, Jun 2006.

\bibitem{Das_2007}
Saurya Das, S.~Shankaranarayanan, and Sourav Sur.
\newblock {Power-law corrections to entanglement entropy of black holes}.
\newblock {\em Phys. Rev. D}, 77:064013, 2008.

\bibitem{Solodukhin:2011gn}
Sergey~N. Solodukhin.
\newblock {Entanglement entropy of black holes}.
\newblock {\em Living Rev. Rel.}, 14:8, 2011.

\bibitem{PhysRevD.102.125025}
S.~Mahesh Chandran and S.~Shankaranarayanan.
\newblock One-to-one correspondence between entanglement mechanics and black hole thermodynamics.
\newblock {\em Phys. Rev. D}, 102:125025, Dec 2020.

\bibitem{PhysRevD.104.085012}
J\'er\^ome Martin and Vincent Vennin.
\newblock Real-space entanglement of quantum fields.
\newblock {\em Phys. Rev. D}, 104:085012, Oct 2021.

\bibitem{Martin:2021qkg}
J.~\'er\^ome Martin and Vincent Vennin.
\newblock {Real-space entanglement in the Cosmic Microwave Background}.
\newblock {\em JCAP}, 10:036, 2021.

\bibitem{Belfiglio:2024qsa}
Alessio Belfiglio, Orlando Luongo, Stefano Mancini, and Sebastiano Tomasi.
\newblock {Entanglement entropy in quantum black holes}.
\newblock 3 2024.

\bibitem{Belfiglio:2023sru}
Alessio Belfiglio, Orlando Luongo, and Stefano Mancini.
\newblock {Entanglement area law violation from field-curvature coupling}.
\newblock 6 2023.

\bibitem{PhysRevD.86.045014}
Vijay Balasubramanian, Michael~B. McDermott, and Mark Van~Raamsdonk.
\newblock Momentum-space entanglement and renormalization in quantum field theory.
\newblock {\em Phys. Rev. D}, 86:045014, Aug 2012.

\bibitem{PhysRevD.102.043529}
Suddhasattwa Brahma, Omar Alaryani, and Robert Brandenberger.
\newblock Entanglement entropy of cosmological perturbations.
\newblock {\em Phys. Rev. D}, 102:043529, Aug 2020.

\bibitem{Brahma:2021mng}
Suddhasattwa Brahma, Arjun Berera, and Jaime Calder\'on-Figueroa.
\newblock {Universal signature of quantum entanglement across cosmological distances}.
\newblock {\em Class. Quant. Grav.}, 39(24):245002, 2022.

\bibitem{PhysRevD.105.123523}
Alessio Belfiglio, Orlando Luongo, and Stefano Mancini.
\newblock Geometric corrections to cosmological entanglement.
\newblock {\em Phys. Rev. D}, 105:123523, Jun 2022.

\bibitem{PhysRevD.107.103512}
Alessio Belfiglio, Orlando Luongo, and Stefano Mancini.
\newblock Inflationary entanglement.
\newblock {\em Phys. Rev. D}, 107:103512, May 2023.

\bibitem{PhysRevD.108.043522}
Suddhasattwa Brahma, Jaime Calder\'on-Figueroa, Moatasem Hassan, and Xuan Mi.
\newblock Momentum-space entanglement entropy in de sitter spacetime.
\newblock {\em Phys. Rev. D}, 108:043522, Aug 2023.

\bibitem{PhysRevD.109.123520}
Alessio Belfiglio, Orlando Luongo, and Stefano Mancini.
\newblock Superhorizon entanglement from inflationary particle production.
\newblock {\em Phys. Rev. D}, 109:123520, Jun 2024.

\bibitem{Boutivas:2023ksg}
Konstantinos Boutivas, Georgios Pastras, and Nikolaos Tetradis.
\newblock {Entanglement and expansion}.
\newblock {\em JHEP}, 05:199, 2023.

\bibitem{Katsinis:2023hqn}
Dimitrios Katsinis, Georgios Pastras, and Nikolaos Tetradis.
\newblock {Entanglement of harmonic systems in squeezed states}.
\newblock {\em JHEP}, 10:039, 2023.

\bibitem{PhysRevD.77.063534}
C.~P. Burgess, R.~Holman, and D.~Hoover.
\newblock Decoherence of inflationary primordial fluctuations.
\newblock {\em Phys. Rev. D}, 77:063534, Mar 2008.

\bibitem{PhysRevD.109.023503}
S.~Mahesh Chandran, Karthik Rajeev, and S.~Shankaranarayanan.
\newblock Real-space quantum-to-classical transition of time dependent background fluctuations.
\newblock {\em Phys. Rev. D}, 109:023503, Jan 2024.

\bibitem{Boutivas:2023mfg}
Konstantinos Boutivas, Dimitrios Katsinis, Georgios Pastras, and Nikolaos Tetradis.
\newblock {Entanglement in cosmology}.
\newblock {\em JCAP}, 04:017, 2024.

\bibitem{PhysRev.56.455}
J.~R. Oppenheimer and H.~Snyder.
\newblock On continued gravitational contraction.
\newblock {\em Phys. Rev.}, 56:455--459, Sep 1939.

\bibitem{Weinberg1972}
Steven Weinberg.
\newblock {\em Gravitation and Cosmology: Principles and Applications of the General Theory of Relativity}.
\newblock John Wiley \& Sons, New York, 1972.

\bibitem{Malafarina2017}
Daniele Malafarina.
\newblock Classical collapse to black holes and quantum bounces: A review.
\newblock {\em Universe}, 3(2):48, 2017.

\bibitem{PhysRevD.101.026016}
Tim Schmitz.
\newblock Towards a quantum oppenheimer-snyder model.
\newblock {\em Phys. Rev. D}, 101:026016, Jan 2020.

\bibitem{PhysRevLett.130.101501}
Jerzy Lewandowski, Yongge Ma, Jinsong Yang, and Cong Zhang.
\newblock Quantum oppenheimer-snyder and swiss cheese models.
\newblock {\em Phys. Rev. Lett.}, 130:101501, Mar 2023.

\bibitem{QOS_plb}
Shi-Hai Dong, Farokhnaz Hosseinifar, Filip Studnička, and Hassan Hassanabadi.
\newblock Some new properties of black holes in the quantum oppenheimer-snyder model.
\newblock {\em Physics Letters B}, 860:139182, 2025.

\bibitem{Bonnor1981}
W.~B. Bonnor and P.~A. Vickers.
\newblock Junction conditions in general relativity.
\newblock {\em General Relativity and Gravitation}, 13(1):29--36, 1981.

\bibitem{misner1973gravitation}
Charles~W. Misner, Kip~S. Thorne, and John~Archibald Wheeler.
\newblock {\em Gravitation}.
\newblock W. H. Freeman and Company, San Francisco, 1973.

\bibitem{chandran2023dynamical}
S.~Mahesh Chandran and S.~Shankaranarayanan.
\newblock Dynamical scaling symmetry and asymptotic quantum correlations for time-dependent scalar fields.
\newblock {\em Physical Review D}, 107(2):025003, 2023.

\bibitem{Leach1977}
P.~G.~L. Leach.
\newblock On a direct method for the determination of an exact invariant for the time-dependent harmonic oscillator.
\newblock {\em The ANZIAM Journal}, 20(1):97--105, 1977.

\bibitem{Lohe2009}
M.~A. Lohe.
\newblock Exact time dependence of solutions to the time-dependent schrödinger equation.
\newblock {\em Journal of Physics A: Mathematical and Theoretical}, 42(3):035307, 2009.

\bibitem{Ermakov2008}
Vasilij~Petrovich Ermakov.
\newblock Second-order differential equations: Conditions of complete integrability.
\newblock {\em Applicable Analysis and Discrete Mathematics}, 2:123--145, 2008.
\newblock Originally published in 1880.

\bibitem{Malafarina:2022wmx}
Daniele Malafarina.
\newblock {Semi-classical dust collapse and regular black holes}.
\newblock 9 2022.

\end{thebibliography}

\newpage

\appendix


\section{Spherical harmonics discretization: procedure in curved FLRW spacetime}\label{appendix:sph_harm_discretization_FLRW}

This appendix outlines the general discretization procedure for a $3+1$-dimensional scalar field theory exhibiting spherical symmetry. Consider the Lagrangian of a massive scalar field in a curved FLRW spacetime, Eq.~\eqref{eq:sf_FLRW_spacetime_Lagrangian}, expressed in spherical coordinates
\begin{widetext}
\begin{equation}\label{eq:scf_lagr_flat_space_spherical_coords_curved}
\begin{aligned}
    2L &=  \int_{\mathcal{D}} \frac{r^2\,\mathrm{d}r}{\sqrt{1-kr^2}}\,\mathrm{d}\Omega \,\, a^3(t) \bigg( \dot{\phi}^2 - \frac{1}{a(t)^2} \bigg[ (1-kr^2)(\partial_r \phi)^2 + \frac{1}{r^2} (\partial_\theta \phi)^2 + \frac{1}{r^2 \sin^2\theta} (\partial_\varphi \phi)^2 \bigg] - \mu^2 \phi^2 \bigg) \\
    &= K + R + A + M,
\end{aligned}
\end{equation}
\end{widetext}
where $\mu$ denotes the field mass, $k$ is the spatial curvature parameter, and the integration domain $\mathcal{D}$ covers the relevant range of the radial coordinate. Here, the Lagrangian is decomposed into four terms: the kinetic term $K$ (involving the time derivative of the field), the radial term $R$ (involving derivatives with respect to the radial coordinate), the angular term $A$ (involving derivatives with respect to the angular variables), and the mass term $M$.

Next, we decompose $\phi$ in terms of real spherical harmonics:
\begin{equation}\label{eq:sph_harm_decomp_0}
    \begin{cases}
        \phi_{lm}(t,r) = \displaystyle \int Y_{lm}(\theta,\varphi) \, \phi(t,r,\theta,\varphi)\, \mathrm{d}\Omega, \\[10pt]
        \phi(t,r,\theta,\varphi) = \displaystyle \sum_{l=0}^{\infty}\sum_{m=-l}^{l} \phi_{lm}(t,r) \, Y_{lm}(\theta,\varphi).
    \end{cases}
\end{equation}
Substituting the decomposition from Eq.~\eqref{eq:sph_harm_decomp_0} into the Lagrangian allows us to evaluate the angular integrals, which simplify due to the orthogonality of the real spherical harmonics. We explicitly address only the integral of the angular term $A$, as it is the most complex. Evaluating the angular integral by parts, we obtain
\begin{equation}
\begin{aligned}
&\int |\nabla \phi|^2_{\text{ang}} \, \mathrm{d}\Omega = -\int \phi (\nabla^2_{\text{ang}} \phi) \, \mathrm{d}\Omega \\
&= -\int \left( \sum_{l m} \phi_{l m} Y_{l m} \right) \left( \sum_{l' m'} \phi_{l' m'} \nabla^2 Y_{l' m'} \right) \mathrm{d}\Omega \\
&= -\sum_{l m} \sum_{l' m'} \phi_{l m}\phi_{l' m'} \int Y_{l m} \left( -\frac{l'(l'+1)}{r^2} Y_{l' m'} \right) \mathrm{d}\Omega \\
&= \sum_{l m} \sum_{l' m'} \phi_{l m} \phi_{l' m'} \frac{l'(l'+1)}{r^2} \int Y_{l m} Y_{l' m'} \, \mathrm{d}\Omega \\
&= \sum_{l m} \frac{l(l+1)}{r^2} \phi_{l m}^2(r).
\end{aligned}
\end{equation}
In the third line, we have utilized the relation $\nabla^2 Y_{lm} = -\frac{l(l+1)}{r^2} Y_{lm}$. Therefore, using the result above, the expression for the angular part of the Lagrangian becomes
\begin{equation}
    \begin{aligned}
        &A=\int_{\mathcal{D}} \frac{r^2\mathrm{d}r }{\sqrt{1 - k r^2}}\, \left[-a|\nabla \phi|^2_{\text{ang}}\right] \\
        &= \sum_{l m} \int_{\mathcal{D}} \frac{r^2\mathrm{d}r }{\sqrt{1 - k r^2}}\, \left[-a\frac{l(l+1)}{r^2}\phi_{lm}^2\right]
    \end{aligned}
\end{equation}
Combining all contributions, the Lagrangian of the field decomposed into spherical harmonics reads
\begin{equation} \label{lagra_spherxp}
    \begin{aligned}
        2L &=K+R+A+M=\\
        &= \sum_{l=0}^{\infty}\sum_{m=-l}^{l} \int_{\mathcal{D}} \frac{r^2\mathrm{d}r }{\sqrt{1 - k r^2}}\, \left[ a^3\dot{\phi}_{lm}^2 \right.\\
        &\left.\qquad - a(1-kr^2)\left(\partial_r \phi_{lm} \right)^2 \right.\\
        &\left.\qquad - a\left(\frac{l(l+1)}{r^2} + (a\mu)^2\right)\phi_{lm}^2 \right].
    \end{aligned}
\end{equation}

Now, we may either perform a change of variables $\tilde{\phi}_{lm} = \frac{r}{(1 - kr^2)^{1/4}} \phi_{lm}$ and subsequently define the conjugate momentum to $\tilde{\phi}_{lm}$, or first define the conjugate momentum to $\phi_{lm}$ and then perform a canonical transformation to eliminate the prefactor $r^2/\sqrt{1 - kr^2}$. Both approaches yield identical results. Here, we adopt the first approach
\begin{equation}
    \begin{aligned}\label{eq:decomposed_lagrangian_tilde_flrw}
        2L &= \sum_{l=0}^{\infty}\sum_{m=-l}^{l} \int_{0}^{\infty} \left[ a^3(\partial_t \tilde{\phi}_{lm})^2 \right. \\
        &\quad \left. -\, a r^2 \sqrt{1 - k r^2} \left[ \frac{\partial}{\partial r} \left( \frac{(1 - k r^2)^{\frac{1}{4}}}{r} \tilde{\phi}_{lm} \right) \right]^2 \right. \\
        &\quad \left. -\, a \left( \frac{l(l + 1)}{r^2} + a^2 \mu^2 \right) \tilde{\phi}_{lm}^2 \right] \, \mathrm{d} r
    \end{aligned}
\end{equation}

Since we are applying a canonical transformation, this ensures the invariance of entanglement entropy when moving from Eq. \eqref{lagra_spherxp} to Eq. \eqref{eq:decomposed_lagrangian_tilde_flrw}. Therefore, we do not need to transform the field back in order to compute the entanglement entropy.

Next, we complete the discretization procedure by redefining the radial variable as $ r_j = b j $, where $ b $ is the ultraviolet cutoff length and $ j $ is a positive integer. The radial discretization scheme can be summarized as:
\begin{equation}
    \begin{aligned}
        &r \rightarrow b j, \qquad \phi_{lm}(r) \rightarrow \phi_{lmj}, \\
        &\frac{\partial \phi_{lm}(r)}{\partial r} \rightarrow \frac{\phi_{lm,j+1} - \phi_{lmj}}{b},\\
        &\int_0^{L} f(r) \, \mathrm{d} r \rightarrow b \sum_{j=1}^{N} f_j.
    \end{aligned}
\end{equation}
Additionally, we employ the midpoint finite difference scheme to approximate terms such as $ f(r) \frac{\partial g(r)}{\partial r} \rightarrow f\left(b\left[j + \frac{1}{2}\right]\right) \frac{g_{j+1} - g_j}{b} $, where $ f(r) $ and $ g(r) $ are generic functions. To obtain the fully discretized Lagrangian we apply the above depicted scheme to Eq.~(\ref{eq:decomposed_lagrangian_tilde_flrw}). We then compute the corresponding discretized Hamiltonian. Furthermore, we can eliminate some $ b $ factors through an additional canonical transformation. The final result is
\begin{equation}
    \begin{aligned}\label{eq:final_discretized_hamiltonian_flrw}
        2bH &= \sum_{l m j} \left[ \frac{\tilde{\pi}^2_{lmj}}{a^3}   +\, a \left(j + \frac{1}{2}\right)^2 \sqrt{1 - k b^2 \left(j + \frac{1}{2}\right)^2} \right. \\
        &\quad \left. \times \left[ \frac{\left(1 - k b^2 (j + 1)^2\right)^{1/4}}{(j + 1)} \tilde{\phi}_{lm(j+1)} - \frac{(1 - k b^2 j^2)^{1/4}}{j} \tilde{\phi}_{lmj} \right]^2 \right. \\
        &\quad \left. +\, a \left( \frac{l(l + 1)}{j^2} + (\mu a b)^2 \right) \tilde{\phi}_{lmj}^2 \right]
    \end{aligned}
\end{equation}
where we have defined the conjugate momentum as
\[ \tilde{\pi}_{lmj} = b a^3\, \partial_t \tilde{\phi}_{lmj}, \]
from the Lagrangian (\ref{eq:decomposed_lagrangian_tilde_flrw}).


\section{Entanglement entropy of a real ground state matrix}\label{appendix:ee_real_cov_matrix}

We aim to compute the entanglement entropy scaling for a Gaussian state characterized by a complex ground state matrix $\boldsymbol{\Sigma}^l$. However, due to the similarities in the calculations, we first demonstrate the computation for a real ground state matrix and subsequently highlight the main complications introduced by the presence of imaginary terms.

Consider the position space expression of a Gaussian state in normal coordinates, which has the form
\begin{equation}
    \Psi = \mathrm{Det}^{\frac{1}{4}}\left(\frac{\boldsymbol{\Sigma}}{\pi}\right) e^{-\frac{1}{2} \boldsymbol{y}^T \boldsymbol{\Sigma} \boldsymbol{y}},
\end{equation}
where $\boldsymbol{\Sigma}$ is a real diagonal matrix. The ground state density matrix is given by
\begin{equation}
    \begin{aligned}
        \rho(\boldsymbol{y},\boldsymbol{y}') &= \langle \boldsymbol{y} | \Psi \rangle \langle \Psi | \boldsymbol{y}' \rangle = \Psi(\boldsymbol{y}) \Psi^*(\boldsymbol{y}') \\
        &= \mathrm{Det}^{\frac{1}{2}}\left(\frac{\boldsymbol{\Sigma}}{\pi}\right) \exp\left\{ -\frac{1}{2} \left( \boldsymbol{y}^T \boldsymbol{\Sigma} \boldsymbol{y} + \boldsymbol{y}'^T \boldsymbol{\Sigma} \boldsymbol{y}' \right) \right\}.
    \end{aligned}
\end{equation}
We use normal coordinates to diagonalize the Hamiltonian, allowing us to express the state in terms of the Hamiltonian's eigenstates and simplifying the calculations \cite{PhysRevD.109.023503}. Next, we aim to trace out $n$ oscillators belonging to a definite spatial region, which we designate as the \virgolette{inside} or inaccessible region. We are working in normal coordinates, which do not correspond to the physical positions of oscillators. Therefore, we revert to the standard spatial coordinates $\boldsymbol{\eta} = \boldsymbol{U}^T \boldsymbol{y}$. We define the vectors $\boldsymbol{\eta}_n^T = (\eta_1, \ldots, \eta_n)$ and $\boldsymbol{\eta}_N^T = (\eta_{n+1}, \ldots, \eta_{N})$, such that $\boldsymbol{\eta}^T = (\boldsymbol{\eta}_n^T, \boldsymbol{\eta}_N^T)$. Tracing out $n$ oscillators yields the out density matrix $\rho_{\text{out}}$ as
\begin{equation}\label{eq:trace_integral}
    \begin{aligned}
        \rho_{\text{out}} &= \mathrm{Tr}_{\text{in}}(\rho) \\
        &= \int \prod_{i=1}^{n} \mathrm{d}\eta_i \, \langle \boldsymbol{\eta}_{n}, \boldsymbol{\eta}_{N} | \Psi \rangle \langle \Psi | \boldsymbol{\eta}_{n}, \boldsymbol{\eta}'_{N} \rangle \\
        &= \mathrm{Det}^{\frac{1}{2}} \left( \frac{\boldsymbol{\Sigma}}{\pi} \right) \int \prod_{i=1}^{n} \mathrm{d}\eta_i \, \exp \left\{ -\frac{1}{2} \left[ \left( \boldsymbol{\eta}_{n}^T, \boldsymbol{\eta}_{N}^T \right) \boldsymbol{\Omega} \begin{pmatrix} \boldsymbol{\eta}_{n} \\ \boldsymbol{\eta}_{N} \end{pmatrix} \right. \right. \\
        &\quad \left. \left. + \left( \boldsymbol{\eta}_{n}^T, (\boldsymbol{\eta}'_{N})^T \right) \boldsymbol{\Omega} \begin{pmatrix} \boldsymbol{\eta}_{n} \\ \boldsymbol{\eta}'_{N} \end{pmatrix} \right] \right\},
    \end{aligned}
\end{equation}
where the trace is performed over the inside region, and we define $\boldsymbol{\Omega} = \boldsymbol{U}^T \boldsymbol{\Sigma} \boldsymbol{U}$, which is a real, symmetric matrix\footnote{This can be easily proven using the facts that $\boldsymbol{\Sigma}$ is diagonal and $\boldsymbol{U}$ is orthonormal: $\boldsymbol{\Omega}^T = (\boldsymbol{U}^T \boldsymbol{\Sigma} \boldsymbol{U})^T = \boldsymbol{U}^T \boldsymbol{\Sigma} \boldsymbol{U} = \boldsymbol{\Omega}$. This also implies $\boldsymbol{A}^T = \boldsymbol{A}$ and $\boldsymbol{D}^T = \boldsymbol{D}$.}.

We now write $\boldsymbol{\Omega}$ in block form as follows:
\begin{equation}
    \boldsymbol{\Omega} =
    \begin{pmatrix}
        \boldsymbol{A} & \boldsymbol{B} \\
        \boldsymbol{B}^T & \boldsymbol{D}
    \end{pmatrix},
\end{equation}
where $\boldsymbol{A} \in \mathbb{R}^{n \times n}$, $\boldsymbol{D} \in \mathbb{R}^{(N-n) \times (N-n)}$, and $\boldsymbol{B} \in \mathbb{R}^{n \times (N-n)}$. The exponent of the integrand in Eq.~\ref{eq:trace_integral} then becomes
\begin{align}
    & -\frac{1}{2} \left( 2 \boldsymbol{\eta}_n^T \boldsymbol{A} \boldsymbol{\eta}_n
    + 2 (\boldsymbol{\eta}_N + \boldsymbol{\eta}_N')^T \boldsymbol{B}^T \boldsymbol{\eta}_n \right. \notag \\
    & \qquad \left. + \boldsymbol{\eta}_N^T \boldsymbol{D} \boldsymbol{\eta}_N
    + \boldsymbol{\eta}_N'^T \boldsymbol{D} \boldsymbol{\eta}_N' \right)
\end{align}
In these calculations, we rely on the fact that each term is a scalar and therefore equal to its transpose, allowing us to group terms. Thus, the integral we need to perform is
\begin{equation}
    \begin{aligned}
        &\int \prod_{i=1}^{n}\mathrm{d}\eta_i \exp\left(-\left(\boldsymbol{\eta}_n^T\boldsymbol{A}\boldsymbol{\eta}_n + \left[\boldsymbol{B}(\boldsymbol{\eta}_N+\boldsymbol{\eta}_N')\right]^T\boldsymbol{\eta}_n\right)\right) \\
        &=\mathrm{Det}^{\frac{1}{2}}\left(\pi\boldsymbol{A}^{-1}\right)\exp\left(\frac{1}{4}(\boldsymbol{\eta}_N+\boldsymbol{\eta}_N')^T\boldsymbol{B}^T\boldsymbol{A}^{-1}\boldsymbol{B}(\boldsymbol{\eta}_N+\boldsymbol{\eta}_N')\right).
    \end{aligned}
\end{equation}
The reduced density matrix of the outside system is then given by
\begin{equation}
    \begin{aligned}
        \rho_{\text{out}} &= \mathrm{Det}^{\frac{1}{2}}\left(\pi \boldsymbol{A}^{-1}\right) \mathrm{Det}^{\frac{1}{2}}\left(\frac{\boldsymbol{\Sigma}}{\pi}\right) \\
        &\quad \times \exp\left\{-\frac{1}{2} \left[ \boldsymbol{\eta}_N^T \boldsymbol{D} \boldsymbol{\eta}_N + \boldsymbol{\eta}_N'^T \boldsymbol{D} \boldsymbol{\eta}_N' \right.\right. \\
        &\quad \left.\left. - \frac{1}{2} (\boldsymbol{\eta}_N + \boldsymbol{\eta}_N')^T \boldsymbol{B}^T \boldsymbol{A}^{-1} \boldsymbol{B} (\boldsymbol{\eta}_N + \boldsymbol{\eta}_N') \right] \right\}.
    \end{aligned}
\end{equation}
We introduce the following definitions
\begin{equation}\label{eq:useful_matrices_def}
        \boldsymbol{\gamma} \coloneqq \boldsymbol{D} - \frac{1}{2}\boldsymbol{B}^T\boldsymbol{A}^{-1}\boldsymbol{B}, \qquad \boldsymbol{\beta} \coloneqq \frac{1}{2}\boldsymbol{B}^T\boldsymbol{A}^{-1}\boldsymbol{B},
\end{equation}
where $\boldsymbol{\beta}^T = \boldsymbol{\beta}$ and $\boldsymbol{\gamma}^T = \boldsymbol{\gamma}$. Using these definitions, we can express the reduced (outside) density matrix as
\begin{align}
    &\rho_{\text{out}}(\boldsymbol{\eta}_N, \boldsymbol{\eta}_N')
    = \mathrm{Det}^{\frac{1}{2}}\left(\pi \boldsymbol{A}^{-1}\right)
    \mathrm{Det}^{\frac{1}{2}}\left(\frac{\boldsymbol{\Sigma}}{\pi}\right) \notag \\
    &\quad \times \exp\left(-\frac{1}{2} \left( \boldsymbol{\eta}_N^T \boldsymbol{\gamma} \boldsymbol{\eta}_N
    + \boldsymbol{\eta}_N'^T \boldsymbol{\gamma} \boldsymbol{\eta}_N'
    - 2 \boldsymbol{\eta}_N'^T \boldsymbol{\beta} \boldsymbol{\eta}_N \right)\right).
\end{align}
We perform two successive coordinate transformations \cite{qft2}. The first transformation diagonalizes $\boldsymbol{\gamma}$, with $\boldsymbol{\eta}_N = \boldsymbol{V} \tilde{\boldsymbol{\eta}}$ and $\boldsymbol{\gamma}_D = \boldsymbol{V}^T \boldsymbol{\gamma} \boldsymbol{V}$. The second change of variable is given by $\bar{\boldsymbol{\eta}} = \sqrt{\boldsymbol{\gamma}_D} \, \tilde{\boldsymbol{\eta}}$. The resulting density matrix for the out system is
\begin{align}
    &\rho_{\text{out}}(\bar{\boldsymbol{\eta}}, \bar{\boldsymbol{\eta}}') = \\
    &=\mathrm{Det}^{\frac{1}{2}}\left(\frac{\boldsymbol{I} - \tilde{\boldsymbol{\beta}}}{\pi}\right) \exp\left(-\frac{1}{2}\left(\bar{\boldsymbol{\eta}}^T \bar{\boldsymbol{\eta}} + (\bar{\boldsymbol{\eta}}')^T \bar{\boldsymbol{\eta}}' - 2\bar{\boldsymbol{\eta}}^T \tilde{\boldsymbol{\beta}} \bar{\boldsymbol{\eta}}'\right)\right),
\end{align}
where
\begin{equation}
    \tilde{\boldsymbol{\beta}} = \boldsymbol{\gamma}_D^{-1/2} \boldsymbol{V}^T \boldsymbol{\beta} \boldsymbol{V} \boldsymbol{\gamma}_D^{-1/2}.
\end{equation}

Finally, we perform the basis change that diagonalizes the remaining non-diagonal matrix $\tilde{\boldsymbol{\beta}}$, namely $\boldsymbol{z} = \boldsymbol{W}\boldsymbol{\bar{\eta}}$, thus giving
\begin{equation}\label{eq:otside_density_matrix_final_form_real}
\begin{aligned}
    &\rho_{\text{out}}(\boldsymbol{z}, \boldsymbol{z}') =  \\
    &=\prod_{i=1}^{N-n}\sqrt{\frac{1 - \tilde{\beta}_i }{\pi}} \exp\left(-\frac{1}{2} \left[ z_i^2 + (z_i')^2 - 2 \tilde{\beta}_i z_i z_i' \right]\right),
\end{aligned}
\end{equation}
where $\tilde{\beta}_i$ is an eigenvalue\footnote{These eigenvalues can be computed more easily as the eigenvalues of $\boldsymbol{\gamma}^{-1}\boldsymbol{\beta}$, which is profitable for numerical implementation, as it reduces computation time.} of $ \tilde{\boldsymbol{\beta}} $. We now aim to compute the von Neumann entropy relative to the state $\rho_{\text{out}}$. Such entanglement entropy is defined by
\begin{equation}
    S(\rho_{\text{out}}) = -\mathrm{Tr}[\rho_{\text{out}}\ln{\rho_{\text{out}}}] = -\sum_k p_k \ln{p_k},
\end{equation}
where $p_i$ are the eigenvalues of $\rho_{\text{out}}$. We might then compute the eigenvalues of the density matrix in Eq. \eqref{eq:otside_density_matrix_final_form_real}. The eigenvalue equation is
\begin{equation}\label{eq:full_eig_probl}
    \int \rho_{\text{out}}(\boldsymbol{z}, \boldsymbol{z}') f_{k}(\boldsymbol{z}') \, \mathrm{d}\boldsymbol{z}' = p_k f_{k}(\boldsymbol{z}).
\end{equation}
Note that the eigenvalue equation is invariant under any non-singular matrix transformation of the $\boldsymbol{z}$ vectors, and thus we can use the out density matrix in the form of Eq.~(\ref{eq:otside_density_matrix_final_form_real}). The reduced density matrix is a product of terms which we denote as $\rho_{\text{out}}^i$. If the eigenfunction can be also expressed as $f_{ki}(\boldsymbol{z}) = \prod_{i=1}^{N-n} f_{ki}(z_i)$, the eigenvalue equation becomes
\begin{equation}
   \prod_{i=1}^{N-n} \int  \rho_{\text{out}}^i(z_i, z_i') f_{ki}(z_i') \, \mathrm{d}z_i' = \prod_{i=1}^{N-n} p_{ki} f_{ki}(z_i),
\end{equation}
where we have introduced $p_{ki}$ such that $p_k = \prod_{i=1}^{N-n} p_{ki}$. Therefore, we can attempt to solve multiple simpler eigenvalue problems
\begin{equation}\label{eq:simpler_eigenv_equation}
   \int  \rho_{\text{out}}^i(z_i, z_i') f_{ki}(z_i') \, \mathrm{d}z_i' = p_{ki} f_{ki}(z_i),
\end{equation}
assuming these have solutions, thereby implicitly solving the full eigenvalue problem Eq.~(\ref{eq:full_eig_probl}). These simpler eigenvalue problems are indeed solvable. Choosing the ansatz
\begin{equation}
    f_{ki}(x) = \exp\left(-\frac{A_i}{2}x^2\right) H_k(B_i x),
\end{equation}
where $H_k(x)$ is the $k$-th Hermite polynomial, the integral can then be evaluated using
\begin{align}\label{eq:hermite_gauss_integral}
    &\int_{-\infty}^{\infty} \exp\left( - (x - y)^2 \right) H_k(\alpha x) \, dx = \\\notag
    &\quad= \sqrt{\pi}(1 - \alpha^2)^{k/2} H_k\left( \frac{\alpha y}{\sqrt{1 - \alpha^2}} \right).
\end{align}
We now impose that our ansatz is indeed an eigenfunction. The only non-degenerate, real solutions with positive $A_i$ are given by
\begin{equation}\notag
    A_i = \sqrt{1 - \tilde{\beta}_i^2}, \qquad B_i = \sqrt{A_i}.
\end{equation}
Substituting the solution back into the result yields
\begin{align}\label{eq:eigenvalue_raw_expression}
    p_{ki} &= \sqrt{\frac{2(1 - \tilde{\beta}_i)}{1 + \sqrt{1 - \tilde{\beta}_i^2}}} \left(1 - \frac{2\sqrt{1 - \tilde{\beta}_i^2}}{1 + \sqrt{1 - \tilde{\beta}_i^2}}\right)^{k/2}, \\
    f_{ki}(z_i) &= \exp\left(-\frac{1}{2}\sqrt{1 - \tilde{\beta}_i^2} \, z_i^2\right) H_k\left((1 - \tilde{\beta}_i^2)^{\frac{1}{4}} z_i \right).
\end{align}
We can then rewrite the eigenvalue (\ref{eq:eigenvalue_raw_expression}) as
\begin{equation}
    p_{ki} = (1 - \xi_i) \xi_i^k \qquad \xi_i = \frac{\tilde{\beta}_i}{1 + \sqrt{1 - \tilde{\beta}_i^2}},
\end{equation}
where $0 \leq \xi_i \leq 1$. We can now compute the von Neumann entropy of the out density matrix, $\rho_{\text{out}}$. Since $\rho_{\text{out}}$ is a product of $\rho_{\text{out}}^i$, due to additivity the total entropy will be the sum of the entropies of each component. We have
\begin{equation}
    \begin{aligned}
        &S(\rho_{\text{out}}^i) = -\sum_{k=1}^{\infty} p_{ki} \ln(p_{ki}) \\
        &= -\sum_{k=1}^{\infty} \left[(1 - \xi_i) \xi_i^k \ln(1 - \xi_i) + k(1 - \xi_i) \xi_i^k \ln(\xi_i)\right] \\
        &= -\left[\ln(1 - \xi_i) + \frac{\xi_i}{1 - \xi_i} \ln(\xi_i)\right],
    \end{aligned}
\end{equation}
where, in the last step, we used known identities. Therefore, the von Neumann entropy of the out density matrix is given by
\begin{equation}\label{eq:ee_in_terms_of_xi}
    S(\rho_{\text{out}}) = -\sum_{i=0}^{N-n} \left[\ln(1 - \xi_i) + \frac{\xi_i}{1 - \xi_i} \ln(\xi_i)\right].
\end{equation}

In order to emphasize the scaling of entanglement entropy with system size, in the main text we wrote $S_n(\rho_{\text{out}}) = S(\rho_{\text{out}})$, thus making explicit the dependence on the number $n$ of traced out oscillators.

\end{document}